\documentclass{article}
\usepackage[utf8]{inputenc}
\usepackage[margin=1in]{geometry}
\usepackage{graphicx} 
\usepackage{subcaption}  

\usepackage{graphicx}%
\usepackage{multirow}%
\usepackage{amsmath,amssymb,amsfonts}%
\usepackage{amsthm}%
\usepackage{mathrsfs}%
\usepackage[title]{appendix}%
\usepackage{xcolor}%
\usepackage{textcomp}%
\usepackage{manyfoot}%
\usepackage{booktabs}%
\usepackage{algorithm}%
\usepackage{algorithmicx}%
\usepackage{algpseudocode}%
\usepackage{listings}%
\usepackage{xcolor}
\usepackage{soul}
\usepackage{makecell}
\usepackage{gensymb}
\usepackage{mathtools}
\usepackage{setspace}
\usepackage{authblk}
\usepackage{wasysym}
\usepackage{physics}
\usepackage{url}

\newcommand{\matr}[1]{\mathbf{#1}}
\newcommand{\bl}[1]{\boldsymbol{#1}}
\newcommand{\Gh}[0]{{\Gamma_{\rm h}}}
\newcommand{\Gg}[0]{{\Gamma_{\rm g}}}

\title{High-Fidelity Simulations of Two Miscible Fluids in Small Scale Turbulent Mixers Using a Variational Multiscale Finite Element Method}
\author[1]{Dongjie Jia\thanks{Correspondence: jia191@purdue.edu}}
\author[1]{Mohammad Majidi}
\author[2,3]{Kurt D. Ristroph}
\author[1]{Arezoo Ardekani\thanks{Correspondence: ardekani@purdue.edu}}
\affil[1]{School of Mechanical Engineering, Purdue University, West Lafayette, Indiana, USA}
\affil[2]{Department of Agricultural and Biological Engineering, Purdue University, West Lafayette, Indiana, USA}
\affil[3]{Davidson School of Chemical Engineering (by courtesy), Purdue University, West Lafayette, Indiana, USA}
\begin{document}

\maketitle

\begin{abstract}
    Turbulent mixers have been widely used in industrial settings for chemical production and increasingly for therapeutic nanoparticle formulation by antisolvent precipitation. 
    The quality of the product is closely related to the fluid and mixing dynamics inside the mixers. 
    Due to the rapid time scales and small sizes of many turbulent mixing geometries, computational fluid dynamics simulations have been the primary tool used to predict and understand fluid behavior within these mixers.
    In this study, we used the residual-based variational multiscale finite element method to perform high-fidelity turbulent simulations on two commonly used turbulent mixers: the multi-inlet vortex mixer (MIVM) and the confined impinging jets mixer (CIJM). 
    We simulated two geometric variations, two-inlets and four-inlets, of the MIVM and two different inflow ratios of the CIJM.
    Through detailed turbulence results, we identify differences in turbulence onset, total energy, and mixing performance of the two MIVM configurations.
    With the CIJM results, we demonstrate the effect of the flow rate ratio on the impingement behavior, and as a result, on the mixing performance and turbulence. 
    The cross-comparison between the two mixers shows key differences in turbulence and mixing behaviors, such as the turbulence onset, the energy decay, and the output mixing index.
    This study demonstrates the importance of a high-accuracy numerical scheme in simulating the turbulent mixers and understanding the similarities and differences among mixers. 
    Furthermore, the results show potential for optimizing the operating conditions to achieve the best mixing performance.
\end{abstract}

\section{Introduction}

There is a major need for fast and effective mixing devices in both laboratory-scale research and in the industrial manufacturing setting to ensure the quality and uniformity of the product when two or more liquid components are combined. 
The fluid dynamic principles of these devices rely on the onset of turbulence to take advantage of chaotic flow to promote rapid mixing. 
The design of these devices is particularly important when chemical reactions or particle formation take place upon mixing.
In these situations, mixing time and uniformity are critical to the quality of the final product. 
In chemical reactions, mixing must be faster than the reaction time to ensure homogeneity of the product. 
Similarly, for nanoparticle production, the mixing time scale needs to be controlled in relation to a certain characteristic particle self-assembly time scale.
To address these requirements, various small-scale mixers have been proposed that use high-energy turbulence to achieve the desired mixing characteristics.

Two of the most commonly used small-scale turbulent mixers are the multi-inlet vortex mixer (MIVM)~\cite{liu2008mixing} and the confined impinging jet mixer (CIJM)~\cite{johnson2003flash,han2012simple}, which have been extensively explored for nanoparticle antisolvent precipitation~\cite{saad2016principles}. 
The use of these mixers to formulate nanoparticles is termed Flash NanoPrecipitation because the mixing they achieve in regular operation is faster than particle self-assembly.
The MIVM injects multiple fluid streams at high speed through two or four tangentially arranged inlets into a circular mixing chamber.
The tangential injection of the streams creates strong turbulence through shear, which, as a result, creates rapid mixing.
The CIJM consists of two opposing inlets, which result in the two fluid streams colliding head-on in an enclosed mixing chamber.
Intense turbulence originates from the impingement point of the two streams and promotes rapid mixing.
For both mixers, performance is dependent on factors including fluid properties, geometric configuration, inlet flow rates, and inlet flow ratio~\cite{markwalter2018design,devos2025impinging}.

Various research studies have attempted to understand the mixers' performance in relation to the flow and mixing characteristics. 
On an industrial level, studies looked at the correlation of operating conditions, i.e., the inlet conditions, and the product yield and quality~\cite{shen2011self,valente2012nanoprecipitation,chow2014assessment}.
These studies treat the fluid dynamics inside the mixers as a ``black box'' that relates the inlets to the products.
To understand the underlying microscale fluid dynamics inside the mixers, several experimental studies have attempted to image the mixing behavior inside the MIVM~\cite{shi2011confocal,liu2017turbulent} and CIJM~\cite{tucker1980mixing,fonte2015flow}. 
However, due to the rapid evolution and three-dimensionality of turbulence, these studies faced challenges in providing complete spatial and temporal information on fluid and mixing dynamics.

In recent decades, computational fluid dynamics (CFD) has emerged as a prominent tool for studying the turbulent mixing dynamics inside these mixers.
Due to numerical stability and cost constraints, some CFD studies of these mixers have used the Reynolds-averaged Navier–Stokes (RANS) model~\cite{liu2008mixing,madadi2023investigation}.
Although the predicted overall mixing performance is comparable to experimental observations, such a model does not capture transient details of the turbulence inside the mixers.
More recently, detached eddy simulations (DES) have been widely used to simulate both the MIVM~\cite{liu2016dynamic,zheng2022preparation,peng2024study} and the CIJM~\cite{metzger2014compartment,metzger2016mixing}.
DES is a hybrid RANS large eddy simulations (LES) model~\cite{marchisio2009large}, where regions of the fluid field that use LES modeling are determined based on RANS predictions. 
Although this technique is a step up from RANS, the resolution and fidelity of the turbulence it captures are still unsatisfactory.
To fully resolve the turbulent nature of the CIJM, Hao et al. performed direct numerical simulation and showed a disk-shaped turbulent structure inside the mixing chamber, which is not observed in the results of the modeling-based simulation studies~\cite{hao2020flow}. 

In this study, with the goal of simulating turbulence with high fidelity, we adopt a different CFD algorithm than those used in existing literature, namely the residual-based variational multiscale (RBVMS) finite element method (FEM), originally proposed by Bazilevs et al.~\cite{bazilevs2007variational}. 
The method is based on the variational multiscale theory of turbulence~\cite{hughes2000large,hughes2001multiscale,holmen2004sensitivity} and the stabilized finite element methods~\cite{brooks1982streamline}.
Unlike RANS or LES, the RBVMS framework does not explicitly model the eddy viscosity.
Instead, the sub-grid scale is calculated based on the point-wise numerical residual for the Navier-Stokes equations.
This form of calculating sub-scale turbulence ensures consistency when the point-wise solution to the Navier-Stokes equations is exact.
The RBVMS method has been extensively used for simulating high-fidelity turbulence in complex geometries~\cite{guerra2013numerical,long2014computation,rasthofer2018recent,jia2021efficient}.
The solution accuracy is close to that of the DNS for various case studies~\cite{bazilevs2010large,bazilevs2007variational}, with significantly lower computational cost.

To the best of our knowledge, this is the first study to simulate either the MIVM or CIJM using the RBVMS method. 
In addition, we are here simulating two miscible liquids with different densities and viscosities – water and ethanol, relevant to the formulation of therapeutic lipid nanoparticles – which requires the solution of the Navier-Stokes equations and the advection diffusion equations simultaneously. 
This study exists both as a performance study to compare the given mixers and as a computational framework for further studies that investigate other types of turbulent mixers. 
This manuscript is organized as follows: we will present the formulation, numerical implementation, and the simulation setup in Section 2. 
We will present the simulation results and compare the mixer performances in Section 3.
We will discuss the results in Section 4 and conclude the study in Section 5.

\section{Methods}
To simulate the mixing behavior of two miscible liquids, we solve the Navier-Stokes equations coupled with the advection-diffusion equation. 
Since the fluids exist only in the liquid phase in the mixers, we consider the fluids to be incompressible.
We also consider the fluids to be Newtonian, as are the liquids simulated in the study. 
The Navier-Stokes equations describe the fluid dynamics and solve for the velocity and pressure.
The advection-diffusion equation describes the mixing behavior of two fluids and solves for the fluid concentration.
We obtain the discretized weak form of the equations using the finite element method and perform time integration using the generalized-alpha method. 
A coupled Newton-Raphson method is employed for iterative root finding for the Navier-Stokes and advection-diffusion equations simultaneously. 
The details of the formulations and numerical schemes used are discussed in the following sections. 

\subsection{Governing equations}

The incompressible Navier-Stokes equations are written on the computational domain, $\Omega$, for all $t \in (0,T)$ as
\begin{equation}
    \begin{alignedat}{3}
    \bl r_M &= \rho \frac{\partial \bl u}{\partial t} + \rho \bl u \cdot \nabla{\bl u} + \nabla p - \nabla\cdot\boldsymbol{\tau} = 0,  \\
    r_C &= \nabla\cdot\bl u  = 0,  \\
    \end{alignedat}
 \label{eqn:NS-og}
\end{equation}
where $\bl u$ is the velocity vector, $p$ is the pressure, and $\rho$ is the mixture density. 
Note that in this study, we neglected the external force term in the Navier-Stokes equations. 
The deviatoric stress tensor, $\boldsymbol{\tau }$, is expressed using the Newtonian constitutive equation as
\begin{equation}
    \boldsymbol {\tau }=\mu \left[\nabla \mathbf{u} +(\nabla \mathbf {u} )^{\intercal }\right],
    \label{eqn:tau_stress}
\end{equation}
where $\mu$ is the mixture dynamic viscosity. 
The Dirichlet and Neumann boundary conditions are represented as
\begin{equation}
    \bl u = \bl g \quad \text{on} \quad \Gg, \quad\quad \bl n\cdot\boldsymbol{\sigma}  = \bl h \quad \text{on} \quad \Gh,
 \label{NS_bc}
\end{equation}
where $\Gg$ and $\Gh$ are the portions of the boundary $\Gamma = \partial \Omega = \Gg \bigcup \Gh$ where a Dirichlet and Neumann boundary condition is imposed, respectively.
$\bl g$ and $\bl h$ are given functions about the boundaries.
The stress tensor, $\boldsymbol{\sigma}$, is defined as $\boldsymbol{\sigma} = -p\bl I+\boldsymbol{\tau}$.

The nonuniform and nonconstant mixture density, $\rho$, and dynamic viscosity, $\mu$, are calculated as
\begin{equation}
    \begin{alignedat}{3}
    \rho &= c\rho_1 + (1-c)\rho_2, \\
    \mu  &= \exp\left[c\ln{\mu_1} + (1-c)\ln{\mu_2}\right],
    \end{alignedat}
 \label{eqn:rho_mu}
\end{equation}
where the subscripts $1$ and $2$ denote the two miscible fluids present in the domain, and $c$ tracks the concentration of fluid 1 in the fluid mixture.
The logarithmic approximation for dynamic viscosity of the mixture is well established by the existing literature~\cite{reed1959viscosities}.
The exact fluid properties used in the study are specified in Section 2.4.

The evolution of the concentration, $c$, is governed by the unsteady advection-diffusion equation, written on the computational domain, $\Omega$, for all $t \in (0,T)$ as
\begin{equation}
    r_{AD}=\frac{\partial c}{\partial t} + \bl u \cdot \nabla{c}  = D\nabla^2c,
\label{eqn:AD-og}
\end{equation}
where $c$ is the volume fraction, and $D$ is the diffusivity of the two liquids in contact. 
The Neumann and Dirichlet boundary conditions for~\eqref{eqn:AD-og} are represented as
\begin{equation}
    c =  g \quad \text{on} \quad \Gg, \quad\quad \bl n\cdot D\nabla c  = h \quad \text{on} \quad \Gh,
 \label{AD_bc}
\end{equation}

\paragraph{Remarks:}
\begin{enumerate}
    \item Since both $\rho$ and $\mu$ are non-uniform and non-constant in our problem, the diffusive term in~\eqref{eqn:NS-og} expands into two terms as $\nabla \cdot \boldsymbol{\tau} = \mu\nabla^2{\bl u} + \nabla\mu \cdot \left[\nabla \mathbf{u} +(\nabla \mathbf {u} )^{\intercal}\right]$. Such expansion is used in this study. 
    \item The notations $\bl r_M$, $r_C$, and $r_{AD}$ indicate the numerical residual for the momentum, continuity, and advection-diffusion equations, respectively, which are needed for the following sections.
\end{enumerate}

\subsection{Finite element method}

In this study, we chose the finite element method to solve the governing non-linear partial differential equations.
The discrete Galerkin form of the governing equations requires a stabilization scheme to mitigate numerical oscillation that occurs in convection/advection-dominant cases and allows the use of equal shape functions for pressure and velocity in solving the Navier-Stokes equations~\cite{brooks1982streamline,hughes1986new, shakib1991new}.
For stabilizing the Navier-Stokes equations, we use the residual-based variational multiscale method (RBVMS). 
The method is formally presented in a series of publications~\cite{bazilevs2007variational,bazilevs2008isogeometric}.
The exact RBVMS formulation used in this study is summarized in the following.

Assuming we have properly defined infinite-dimensional trial solution and test function spaces, $\mathscr{V}$ and $\mathscr{W}$, respectively, the weak statement of the Navier-Stokes equations is given as: find $\left\{\bl u,p \right\}\in \mathscr{V}$ such that for all $\left\{\bl w,q \right\}\in \mathscr{W}$,
\begin{equation}
\begin{split}
 \mathit{B}\bigg(\left\{\bl u,p \right\}, \left\{\bl w,q \right\}\bigg) = \bigg(\bl w,\rho \frac{\partial \bl u}{\partial t}\bigg)_\Omega &+ \bigg(\bl w,\rho \bl u \cdot \nabla{\bl u}\bigg)_\Omega  + \bigg(q,\nabla\cdot\bl u\bigg)_\Omega  - \bigg(\nabla\bl w, p\bigg)_\Omega \\
 &+ \bigg(\nabla \bl w,\mu\nabla\bl u\bigg)_\Omega-\bigg(\bl w,\nabla\mu \cdot \left[\nabla \mathbf{u} +(\nabla \mathbf {u} )^{\intercal }\right]\bigg)_\Omega = 0.
 \end{split}
 \label{NS_weak}
\end{equation}
The inner product notation for two vector functions $\bl f(\bl x)$ and $\bl g(\bl x)$ over $S$ is defined as
\begin{equation}
    \bigg(\bl f, \bl g\bigg)_S \coloneqq \int_S \bl f^{\intercal} \bl g \dd S.
    \label{inner_def}
\end{equation} 

From the weak form, the RBVMS method performs a multi-scale decomposition of the form $\mathscr{V} = \mathscr{V}^h+\mathscr{V}'$, where $\mathscr{V}^h$ denotes the finite dimensional space associated with the finite element discretization and $\mathscr{V}'$ denotes the subgrid scale.
Using this definition, we decompose the velocity and pressure as $\left\{\bl u,p \right\} = \left\{\bl u^h,p^h \right\} + \left\{\bl u',p' \right\}$, where $\left\{\bl u^h,p^h \right\} \in \mathscr{V}^h$ and $\left\{\bl u',p' \right\} \in \mathscr{V}'$. 
The subgrid-scales $\left\{\bl u',p' \right\}$ are modeled based on the residual of the Navier-Stokes momentum ($\bl r_M$) and continuity ($r_C$) equations, respectively.
Incorporating this form with~\eqref{NS_weak}, the final discrete weak form of the RBVMS formulation is given as
\begin{equation}
\begin{split}
 \mathit{B}\bigg(\left\{\bl u^h,p^h \right\}, \left\{\bl w^h,q^h \right\}\bigg) + \mathit{B}_{mod}\bigg(\left\{\bl u^h,p^h \right\}, \left\{\bl w^h,q^h \right\}\bigg)= \bigg(\bl w^h,\bl h\bigg)_\Gh,
 \end{split}
 \label{NS_RBVMS}
\end{equation}
where $\mathit{B}_{mod}$ contains all terms from modeling the subgrid scales as
\begin{equation}
\begin{split}
\mathit{B}_{mod}\bigg(\left\{\bl u^h,p^h \right\}, \left\{\bl w^h,q^h \right\}\bigg)= \sum_{e=1}^{n_{el}}\bigg(\left(\rho^h \bl u^h\cdot\nabla\bl w^h + \nabla q\right),\tau_M\bl r_M\left(\bl u^h,p^h\right)\bigg)_{\Omega_e} + \sum_{e=1}^{n_{el}}\bigg(\nabla\cdot \bl w^h,\tau_C r_C\left(\bl u^h\right)\bigg)_{\Omega_e} \\
+\sum_{e=1}^{n_{el}}\bigg(\rho^h \bl w^h,\tau_M \bl r_M\left(\bl u^h,p^h\right)\cdot\nabla\bl u^h\bigg)_{\Omega_e} - \sum_{e=1}^{n_{el}}\bigg(\rho^h \nabla \bl w^h,\tau_M\bl r_M\left(\bl u^h,p^h\right)\otimes\tau_M\bl r_M\left(\bl u^h,p^h\right)\bigg)_{\Omega_e} \\
+ \sum_{e=1}^{n_{el}}\bigg(\tau_M\bl r_M\left(\bl u^h,p^h\right)\cdot\nabla\bl w^h,\bar{\tau}\tau_M\bl r_M\left(\bl u^h,p^h\right)\cdot\nabla\bl u^h\bigg)_{\Omega_e}.
\end{split}
 \label{NS_mod}
\end{equation}
The stabilization parameters $\tau_M$, $\tau_C$ and $\bar{\tau}$ are calculated at each element as
\begin{equation}
    \begin{alignedat}{1}
        \tau_M &= \left(\omega^2+\matr{u}^h \cdot \boldsymbol{\xi} \matr{u}^h + C_{\mathrm{I}} \nu^2\boldsymbol{\xi}:\boldsymbol{\xi}\right)^{-1/2},\\
        \tau_C &= \left(\tau_M \mathrm{tr}\left(\boldsymbol{\xi}\right)\right), \\
        \bar{\tau} &= \left(\tau_M\bl r_M\cdot\boldsymbol{\xi}\tau_M\bl r_M\right)^{-1/2},
    \end{alignedat}
    \label{tau}
\end{equation}
where $\nu$ is the kinematic viscosity, $\boldsymbol{\xi}$ is the covariant tensor obtained from the physical-parent elements mapping, and the shape-function-dependent constant $C_{\mathrm{I}} = 3$. 
$\omega$ is the fluid domain natural frequency calculated from the flow acceleration and velocity as $\omega = \frac{\lVert\frac{\partial{\bl u^h}}{\partial t}\rVert}{\lVert\bl u^h\rVert}$, where $\lVert\cdot\rVert$ denotes the $l^2$-norm~\cite{jia2023time}. 

For the advection-diffusion equation, we used the Streamline upwind Petrov–Galerkin (SUPG) method to stabilize the Galerkin's weak form of the equations. 
The stabilized discrete weak form is given as finding $c^h$ such that for all $w^h$, 
\begin{equation}
    \bigg(w^h,\frac{\partial c^h}{\partial t}\bigg)_{\Omega} + \bigg(w^h,\bl u^h \cdot \nabla{c^h}\bigg)_{\Omega} + \bigg(\nabla w^h,D\nabla c^h\bigg)_{\Omega} + \bigg(\bl u^h\cdot \nabla w^h,\tau_{SUPG}r_{AD}\left(\bl u^h,c^h\right)\bigg)_\Omega=\bigg(w^h,h\bigg)_\Gh.
    \label{AD_SUPG}
\end{equation}
The stabilization parameter, $\tau_{\mathrm{SUPG}}$, is calculated similar to that of $\tau_M$ as
\begin{equation}
     \tau_{\mathrm{SUPG}} = \left(C_{S}\omega_c^2+\matr{u}^h \cdot \boldsymbol{\xi} \matr{u}^h + C_{\mathrm{I}} D^2\boldsymbol{\xi}:\boldsymbol{\xi}\right)^{-1/2},
     \label{tau_SUPG}
\end{equation}
where $\omega_c = \frac{\lVert\frac{\partial{c^h}}{\partial t}\rVert}{\lVert c^h\rVert}$.
Different from the Navier-Stokes stabilization, a stabilization constant, $C_S=10^3$ is added to the frequency term for improved stability to account for the sharp changes in the concentration solution. 

A backflow stabilization scheme is required to prevent numerical instabilities caused by reversal flow through $\Gh$, where a Neumann boundary condition is imposed~\cite{Esmaily2011backflow}. 
This scheme is especially critical for highly turbulent and low mean velocity flows, which can cause significant elemental backflow.
The stabilization is achieved by adding
\begin{equation}
    \bigg(\bl w^h, \frac{\rho^h}{2}\beta  |\bl u^h\cdot\bl n|_{-} \bl u^h \bigg)_\Gh, 
\end{equation}
to the right-hand side of \eqref{NS_RBVMS}, where 
\begin{equation*}
     |\bl u\cdot \bl n|_- = \frac{\bl u\cdot \bl n - |\bl u\cdot \bl n|}{2},
\end{equation*}
and $\beta \in [0, 1]$ is a user-defined coefficient, which we used $\beta = 0.2$ in all cases.

\paragraph{Remarks:}
\begin{enumerate}
    \item The use of the subgrid scale term completely circumvents the use of eddy viscosity, which is an \textit{ad hoc} mechanism in LES. The detailed derivation of the subgrid scale terms is previously published in~\cite{bazilevs2007variational,bazilevs2008isogeometric}.
    \item The stabilization parameter $\tau$ was historically developed within the theory of stabilized methods. In this study, we used a recently proposed form of $\tau$ that uses a frequency term, $\omega$ for the unsteady stabilization~\cite{esmaily2023stabilized,jia2023time,esmaily2024new,jia2025introducing,esmaily2025augmented}. Alternative forms of $\tau$ are proposed in~\cite{hsu2010improving,balu2023direct,takizawa2023variational}.  
\end{enumerate}

\subsection{Numerical implementation}
The above finite element formulations are implemented in an in-house solver named Multiphysics Finite Element Solver (MUPFES)~\cite{moghadam2013modular,esmaily2015bi,esmaily2013new}. 
The solver has been extensively validated and deployed to simulate complex turbulent flows in various studies~\cite{jia2022characterization,jia2021simulation,rydquist2024investigating,mansour2025multi}. 
The solver uses a second-order generalized-$\alpha$ method for the time integration~\cite{jansen2000generalized}, where $\rho_{\infty} = 0.2$ for all simulations.
Linear shape functions are used for all elements. 

The nonlinear systems of ODE in~\eqref{NS_RBVMS} and~\eqref{AD_SUPG} are linearized and iteratively solved using a modified Newton-Raphson method.
The two equations are coupled through density, $\rho$, and viscosity, $\mu$, in the Navier-Stokes equations and advective velocity, $\bl u^h$, in the advection-diffusion equation. 
The equations are solved in a weakly coupled manner, meaning the linear systems for the Navier-Stokes equations and the advection-diffusion equations are solved separately. 
The variables are coupled through the construction of the linear system at each Newton-Raphson iteration, given as solving
\begin{equation}
\begin{alignedat}{1}
    \bl K^{(n)} \bl Y^{(n+1)} &= -\bl R^{(n)}, \\
    \bl K_{AD}^{(n)} \Delta\dot{\bl C}^{(n+1)} &= -\bl R_{AD}^{(n)},
\end{alignedat}
\end{equation}
where $n$ indicates the number of Newton-Raphson iterations. 
$\bl Y = [\Delta\dot{\bl U}, \Delta\bl P]^{\intercal}$ and $\Delta \dot{\bl C}$ are the corrections to the unknown variable vectors that contain all nodal values of time derivative of velocity and pressure, and time derivative of concentration, respectively.
$\bl R = [\bl R_M, \bl R_C]^\intercal$ and 
$\bl R_{AD}$ are the nodal residual vectors for the momentum equation, continuity equation, and advection-diffusion equation, respectively.
$K$ and $K_{AD}$ are the tangent matrices (or stiffness matrices), which are defined as
\begin{equation}
\bl K = \begin{bmatrix}
    \frac{\partial\bl R_M}{\partial\dot{\bl U}} & \frac{\partial\bl R_M}{\partial\bl P} \\
    \frac{\partial\bl R_C}{\partial\dot{\bl U}} & \frac{\partial\bl R_C}{\partial\bl P}
\end{bmatrix},\; \bl K_{AD}=\frac{\partial\bl R_{AD}}{\partial\dot{\bl C}}.
\end{equation}
The exact terms of the tangent matrices, omitted here for brevity, can be found in existing literature~\cite{bazilevs2008isogeometric,esmaily2013new}.

The Navier-Stokes equations and the advection-diffusion equation are calculated in a two-way coupled manner.
For the Navier-Stokes equations, since the fluid density and viscosity are calculated based on the concentration, as defined in~\eqref{eqn:rho_mu_bound}, the tangent and residual terms are constructed as $\bl K^{(n)}\left(\bl u^{(n)},p^{(n)},c^{(n)}\right)$ and $\bl R^{(n)}\left(\bl u^{(n)},p^{(n)},c^{(n)}\right)$, respectively. 
Since the linear system for the Navier-Stokes equation is solved first at each Newton-Raphson iteration, the velocity is updated to $\bl u^{(n+1)}$ when constructing the linear system for the advection-diffusion equation. 
Therefore, the tangent and residual matrices for the advection-diffusion equation are computed as $\bl K_{AD}^{(n)}\left(\bl u^{(n+1)}\right)$ and $\bl R_{AD}^{(n)}\left(\bl u^{(n+1)},c^{(n)}\right)$, respectively. 

Due to the continuous nature of the Galerkin's method, the concentration, $c$, which is physically a bounded function on $[0,1]$, can have a small numerical overshoot or undershoot at certain locations, such as near the fluid's interface.
In the numerical implementation, we impose the physical bounds for the concentration when calculating density and viscosity as
\begin{equation}
    \begin{alignedat}{3}
    c^* &= \max(\min(c^h,1),0)\\
    \rho^h &= c^*\rho_1 + (1-c^*)\rho_2, \\
    \mu^h  &= \exp{\left[c^*\ln{\mu_1} + (1-c^*)\ln{\mu_2}\right]}, \\
    \end{alignedat}
 \label{eqn:rho_mu_bound}
\end{equation}
Note that this bound is not imposed on the unknown variable $c$ itself. 
Therefore, it will not violate the conservation law described by the advection-diffusion equation. Moreover, this implementation has minimal effect on the mass conservation of the fluids, which is detailed in Appendix~\ref{app:mass}.

The linear system for the Navier-Stokes equations is solved using a bi-partitioned iterative algorithm specifically designed for solving the linear system from the Navier-Stokes equations~\cite{esmaily2015bi}.
The linear system for the advection-diffusion equation is solved using a preconditioned generalized minimal residual (GMRES) algorithm~\cite{shakib1989multi}.
The tolerances are set to 0.4 and 0.1 for the Navier-Stokes linear solver and GMRES, respectively.
At each time step, the Newton-Raphson iteration converges after the residual is reduced by four orders of magnitude. 

\subsection{Simulation setup}
In all simulations in this study, the two liquids under mixing are ethanol and water. 
The density and dynamic viscosity of ethanol are taken at room temperature as $\rho_1=0.789\;\mathrm{g/cm^3}$ and $\mu_1 = 1.2 \;\mathrm{cP}$, respectively.
The density and dynamic viscosity of water are taken at room temperature as $\rho_2=1\;\mathrm{g/cm^3}$ and $\mu_2 = 1 \;\mathrm{cP}$, respectively.
The subscripts 1 and 2 are distinguished in equation~\eqref{eqn:rho_mu}.
The diffusivity of ethanol in water $D = 1.23\times10^{-5}\;\mathrm{cm^2/s}$, used in equation~\eqref{eqn:AD-og}, is taken from literature~\cite{hao1996binary,hills2011diffusion}.

Four CFD simulations are performed: namely, MIVM two-inlet and MIVM four-inlet, both with ethanol-water flow rate ratios of 1:1; and CIJM~\cite{stephandevelopment2022,helix2025} with two different ethanol-water flow rate ratios of 1:3 and 3:3, which have been reported by companies marketing impinging jets mixers. 
The geometry and mesh of these mixers are detailed in the following sections.
For the Navier-Stokes solver, Dirichlet conditions are imposed with flat profiles, implemented as $\bl g = (Q/A)\bl n$ in equation~\ref{NS_bc}, where $Q$ is the inlet flow rate and $A$ is the inlet area. 
Zero Neumann conditions ($\bl h=0$) are imposed on the outlets and no-slip conditions are imposed on the walls ($\bl g=0$).
For the advection diffusion equation, Dirichlet conditions are imposed on the ethanol inlets ($g=1$) and water inlets ($g=0$).
Zero Neumann conditions are imposed on the walls and outlets ($h=0$).

For all mixer geometries studied, the inlets are extended by 0.2 cm (MIVM) and 0.1 cm (CIJM) to allow the flows to fully develop in the inlet channels.
The outlets are extended by the same amounts to account for the numerical backflow stabilization scheme.
These extended outlet regions will be discarded when processing the simulation results.
All simulations are run in parallel using 500 cores at 2.2GHz on Purdue University's Negishi cluster with AMD Epyc ``Milan" processors. 
Each simulation takes around 10 hours to reach statistical convergence.
The time step sizes for the MIVM and CIJM simulations are $10^{-5}$ and $5\times10^{-5}$ seconds, respectively.
All simulations are run for 4000 time steps to reach statistical convergence. 
The solutions are then saved for an additional 1000 time steps as used as converged results. 

\subsubsection{Mixer Geometry}
Three mixer geometries are used in this study: a two-inlet MIVM, a four-inlet MIVM, and a CIJM. The MIVM geometries include rectangular inlet channels, a disk-shaped mixing chamber, and a cylindrical outlet channel. The CIJM has two opposing cylindrical inlet channels and a cylindrical mixing chamber with a cone-shaped tip. Table~\ref{tab:mixer_geometry} summarizes the dimensions of each mixer. 
For both the two-inlet and four-inlet MIVM geometries, the measurements are identical except for the number of inlet channels. 
For the CIJM, the inlets are located at 0.3 cm above the mixing chamber outlet. 

 \begin{table}[h!]
\centering
\caption{Geometrical dimensions of the MIVM and CIJM components. All values are in centimeters. The measurements are without the extended inlets and outlets. }
\begin{tabular}{llcc}
\hline
\textbf{Unit: cm} & \textbf{MIVM} & \textbf{CIJM} \\
\hline
Inlet cross-section & $0.113W\times0.145H$ & \text{\diameter}0.05 \\
Inlet length        & 0.652 & 0.305 \\
Chamber diameter    & 0.593 & 0.249 \\
Chamber height      & 0.145 & 0.481 \\
Outlet diameter     & 0.157 & --- \\
Outlet length       & 0.858 & --- \\
Tip height          & ---   & 0.075 \\
\hline
\end{tabular}
\label{tab:mixer_geometry}
\end{table}

\begin{figure}[h]
    \centering
    \begin{subfigure}[b]{0.3\textwidth}
        \includegraphics[width=\textwidth , trim={0cm 0cm 0cm 0cm},clip]{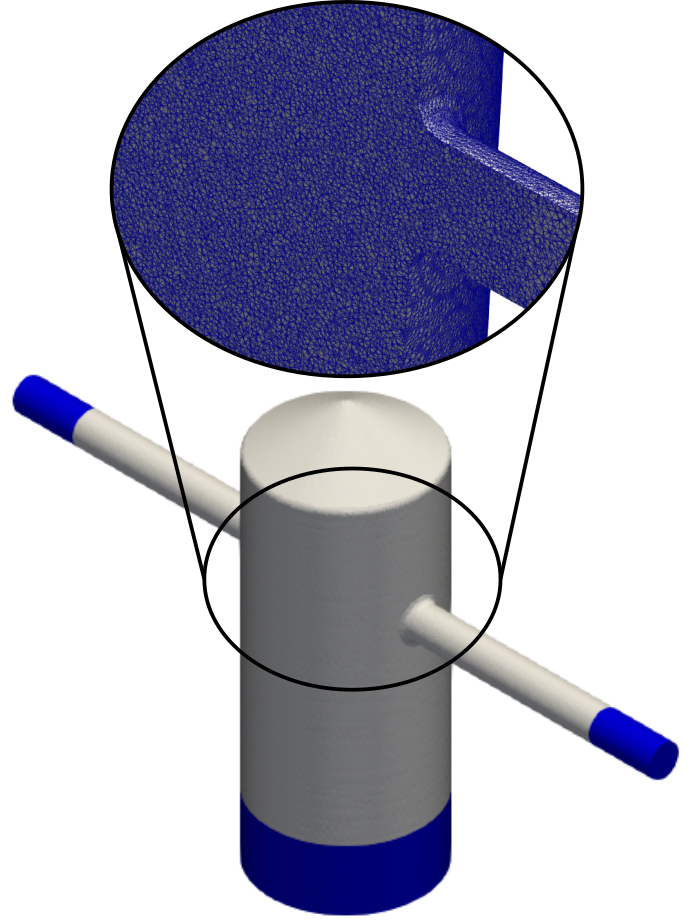}
        \caption{ }
    \end{subfigure}
        \begin{subfigure}[b]{0.3\textwidth}
        \includegraphics[width=\textwidth , trim={0cm 0cm 0cm 0cm},clip]{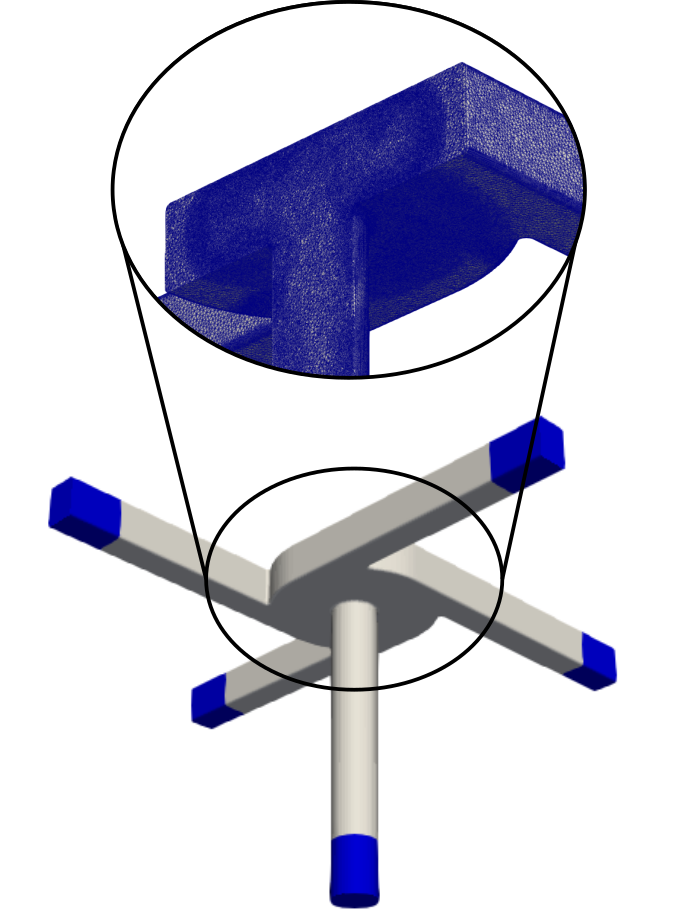}
        \caption{ }
    \end{subfigure}
            \begin{subfigure}[b]{0.3\textwidth}
        \includegraphics[width=\textwidth, trim={0cm 0cm 0cm 0cm},clip]{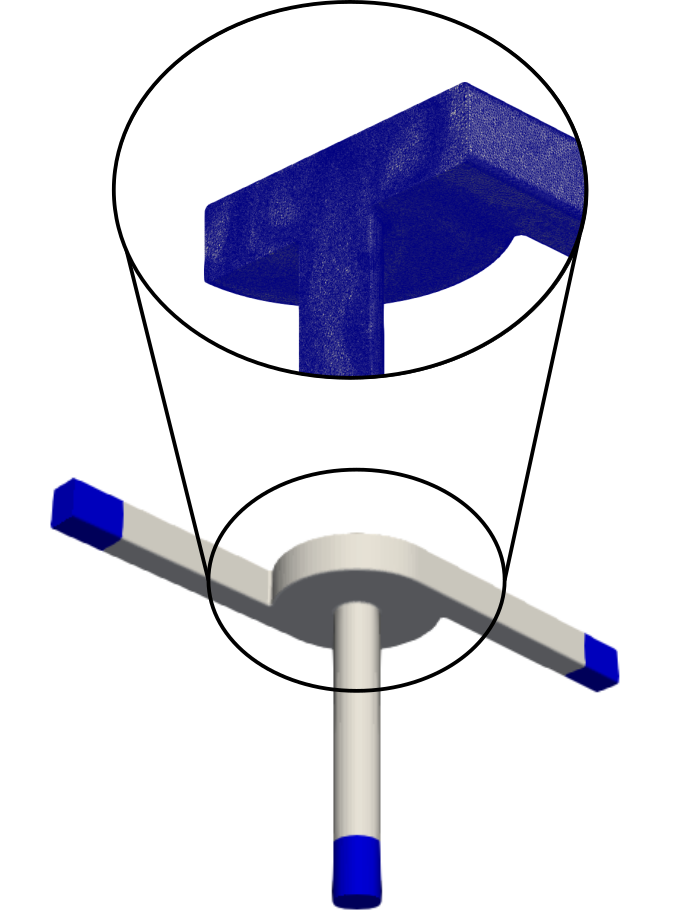}
        \caption{ }
    \end{subfigure}
    \caption{Geometries of (a) the CIJM, (b) the four-inlet MIVM, and (c) the two-inlet MIVM with zoomed-in cross-sectional view of the mesh. Note that the sizes of the CIJM and MIVM relative to one another are not to scale. }
\label{fig:cij_mesh}
\end{figure}

Figure~\ref{fig:cij_mesh} shows the isometric views of the mixers used in this study, with a zoomed-in cross-sectional view of the mesh. 
The mesh generation technique will be discussed in the next section. 
The extended inlets for flow development and the extended outlets for backflow stabilization are marked in blue.
All geometries are created using Autodesk Fusion~\cite{fusion360}, a Computer-Aided Design (CAD) software. 

\subsubsection{Mesh construction}

Due to the shear-dominant nature of MIVM flow, the concentration field exhibits sharp changes at locations where the two fluid streams are in parallel, such as near the stream inlets in the mixing chamber.
At these locations, the spatial discretization requires sufficiently high resolution to limit the numerical error.
In this study, to ensure the solution accuracy and keep the computational cost reasonable, we implement a physics-based iterative mesh refinement algorithm based on the previous iteration simulation results and an \textit{a priori} error estimate. 
For linear shape functions, the numerical solution error, $E$, scales with the element size, $h$, and the concentration, $c$, by
\begin{equation}
E \propto h^2 \left\| \nabla^2 c \right\|.
\end{equation}
To limit the local error, a smaller element size is required in regions where the second derivative of the concentration is large.
 
The iterative mesh generation scheme is implemented as follows. 
Initially, a coarse mesh consisting of approximately two million uniform tetrahedral elements is generated using \textsc{TetGen}~\cite{hang2015tetgen}. 
This initial mesh is used to perform the CFD simulations until statistical convergence. 
The converged solution is saved every five time steps for a total duration of 1000 time steps.
Using the series of saved results, the target element edge length for the next iteration, $h_a$, at mesh node $a$ is defined as 

\begin{equation}
h_a \coloneqq \min_{t\in T} \sqrt{C_r \frac{\lVert\bar{c}\left(t\right)\rVert}{\lVert\nabla^2 c_a\left(t\right)\rVert}},
\end{equation}
where $T$ is the total solution save period, $C_r = 10^{-3}$ is a user-defined non-dimensional constant that relates to the acceptable error level, $\bar{c}$ is the average concentration value in the fluid domain, and $\nabla^2 c_a$ is the Laplacian of the concentration. 
The resulting list of target edge length at each node is then passed to \textsc{TetGen} to perform the remesh.
Two iterations are used to generate the final mesh.
This procedure adds elements in regions of sharp changes in the concentration field to reduce overall error and improve the numerical stability of the solution. 
The resulting two-inlet MIVM mesh contains 12{,}152{,}451 elements and the four-inlet MIVM mesh contains 15{,}169{,}618 elements.
A sectional view of the final mesh is shown in Figure~\ref{fig:cij_mesh}.

The iterative mesh generation scheme is not used for the CIJM geometry. 
This is due to the pure chaotic nature of the CIJM impinging flow, where the fluid-fluid interface is constantly shifting in time and space.
Instead, we used a multi-resolution approach to create the mesh.
The inlet tubes and the outflow stabilization region are meshed with elements with an edge length of $4.4\times 10^{-3}$ cm, while the center chamber is meshed with eight times the resolution using element edge length of $2.2\times 10^{-3}$ cm. 
The resulting mesh contains 12{,}977{,}911 tetrahedral elements. 
A cross-sectional view of this mesh is shown in Figure~\ref{fig:cij_mesh}.

\section{Results} 

All MIVM simulations are performed at the total flow rate of $80\,\text{mL/min}$. In the four-inlet case, each inlet operates at $20\,\text{mL/min}$, while in the two-inlet case, each inlet operates at $40\,\text{mL/min}$. 
The arrangement of the inlet streams in the four-inlet case is symmetric: two opposing inlets supply ethanol, and the other two opposing inlets supply water. 
To calculate the inlet Reynolds number, we used the mean inlet velocity as the characteristic velocity and the hydraulic diameter of the inlet channel as the characteristic length. 
The Reynolds numbers for the four-inlet MIVM are 170 and 259 for the ethanol and water streams, respectively.
For the two-inlet MIVM, the Reynolds numbers are doubled, at 340 and 518 for ethanol and water, respectively.

For the CIJM, two different inlet flow-rate ratios (FRRs) are examined: three-to-three and one-to-three (ethanol-to-water). 
For the three-to-three case, both the ethanol and the water inlet flow rates are $60\,\text{mL/min}$, resulting in a total flow rate of $120\,\text{mL/min}$. 
The resulting Reynolds numbers are 1674 and 2546 for the ethanol and water streams, respectively. 
For the one-to-three case, the ethanol flow rate is reduced to $20\,\text{mL/min}$ while the water flow rate is kept at $60\,\text{mL/min}$, yielding a lower total flow rate of $80\,\text{mL/min}$. 
The Reynolds numbers are 558 and 2546 for the ethanol and water streams, respectively. 

\subsection{Multi-inlet vortex mixers}
 
\begin{figure}[h!]
    \centering
        \includegraphics[clip,width=\textwidth]{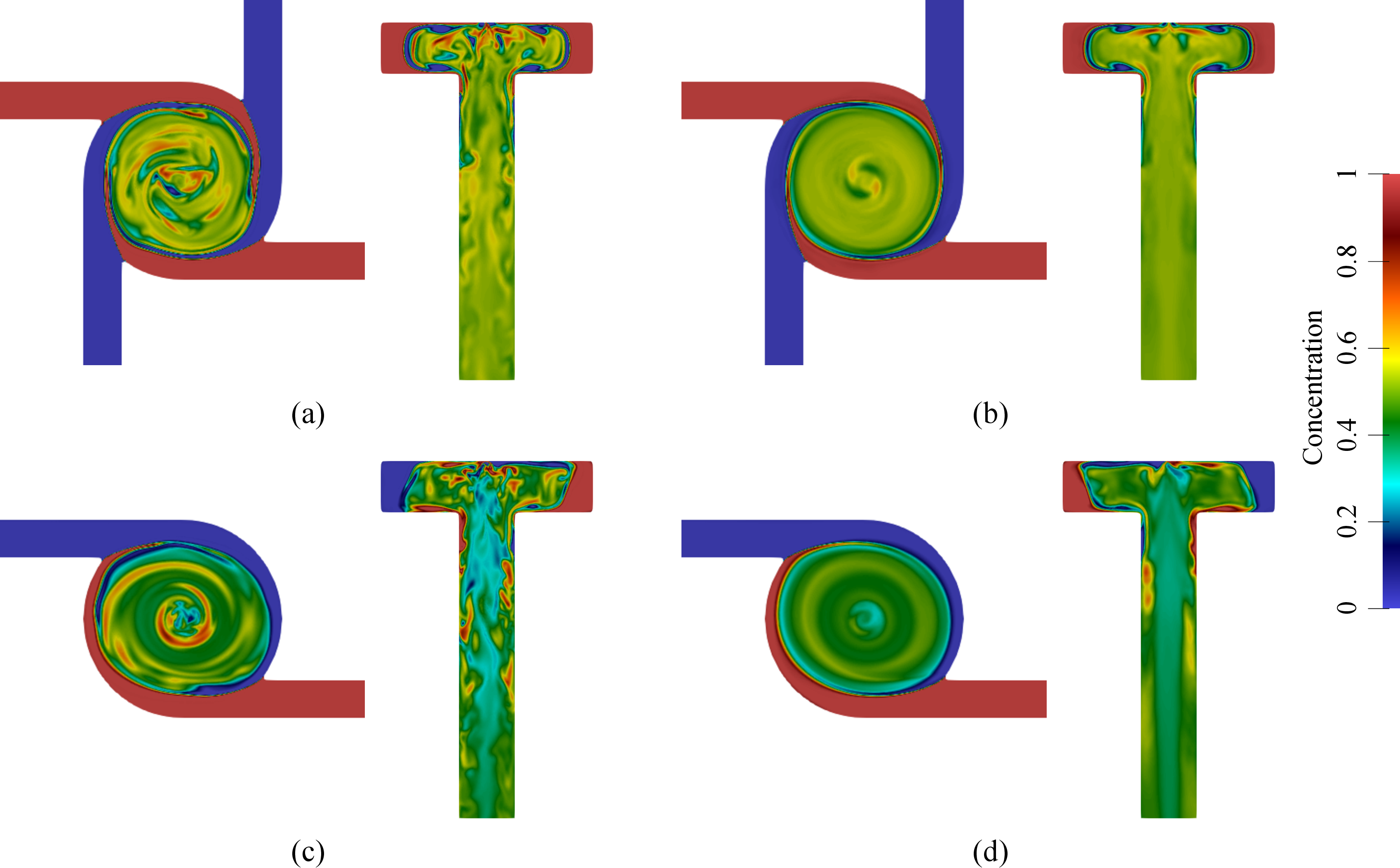}
 \caption{Comparison of concentration fields for four-inlet (top row) and two-inlet (bottom row) MIVM, both operating at 80 mL/min total flow rate. (a) and (c) are instantaneous snapshots, and (b) and (d) are averaged over time.}
    \label{fig:concentration_mivm}
\end{figure}

Figure~\ref{fig:concentration_mivm} shows both instantaneous and time-averaged ethanol concentration for the four-inlet and two-inlet MIVM mixers. 
The top view is taken at the mid-plane of the mixing chamber, and the side view is taken at the center-plane of the outlet channel. 
The four-inlet case shows noticeably faster mixing and a more uniform concentration field towards the outlet.
Due to more contact area between the two fluids created by the four-inlet configuration, the fluids are effectively mixed earlier in the mixing chamber, evident from the large area that is close to 0.5 in concentration (yellow-green color) from the top view.

To provide a quantitative comparison of the mixing performance across different cases, we define the mixing index, $MI$, as
\[
MI \coloneqq 1 - \sqrt{\frac{\sigma^2}{\sigma^2_{\text{max}}}},
\]
where $\sigma^2$ is the variance and $\sigma^2_{\text{max}}$ is the maximum possible variance. 
The mixing index value ranges from 0 (separated fluids) to 1 (homogeneous fluids). 
In this study, we define the variance, $\sigma^2$, based on the concentration, $c$, as
\begin{equation}
    \sigma^2\coloneqq\frac{\int_S (c - \bar{c})^2\dd S}{\int_S \dd S}
    \label{eqn:variance}
\end{equation}
where $\bar{c} = \int_Sc \; \dd S/\int_S\dd S$ and $\sigma_{\mathrm{max}}^2 = 0.25$.
By this definition, a mixing index value close to 1 indicates ideal mixing between water and ethanol.

\begin{figure}[H]
    \centering
    \begin{subfigure}[b]{0.47\textwidth}
        \includegraphics[width=\textwidth]{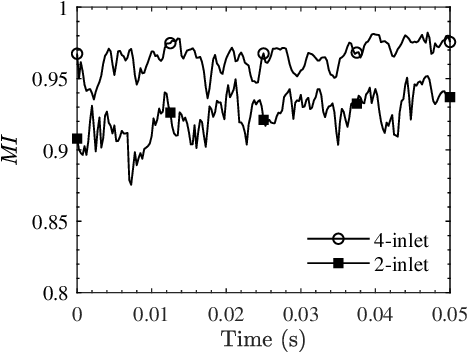}
        \caption{ }
    \end{subfigure}
    \hfill
    \begin{subfigure}[b]{0.464\textwidth}
        \includegraphics[width=\textwidth]{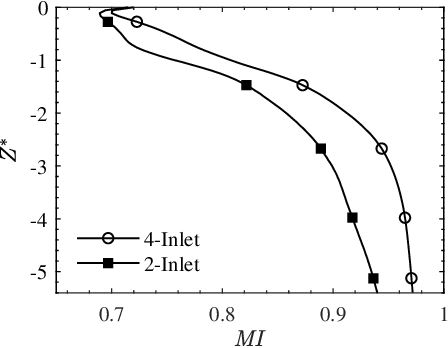}
        \caption{ }
    \end{subfigure}
    \caption{Comparison of mixing index between two-inlet and four-inlet MIVMs. (a) Temporal evolution of the mixing index ($MI$) measured at the outlet exit. (b) Average mixing index evolution along the outlet direction ($Z^*$). }
    \label{fig:MIcomparison_mivm}
\end{figure}

Figure~\ref{fig:MIcomparison_mivm} compares the mixing index between the two-inlet and four-inlet MIVM cases. 
Figure~\ref{fig:MIcomparison_mivm}a shows the temporal evolution of the mixing index value at the outlet exit circular cross-section. 
The four-inlet case shows higher mixing index values with less fluctuation, indicating more complete mixing and a stable mixture at the outlet compared to the two-inlet case. 
Figure~\ref{fig:MIcomparison_mivm}b shows the evolution of the mixing index along the outlet channel. 
The vertical position is non-dimensionalized using the outlet diameter, $d$, as $Z^*=Z/d$.
$Z^*=0$ is the beginning of the outlet channel connected to the mixing chamber. 
The mixing index is calculated over the circular cross-section at each $Z^*$ and averaged over time.
As shown in the figure, the four-inlet mixer reaches a mixing index above 0.9 at around $Z^* = -1$, while the two-inlet mixer does not achieve the same level of mixing index until close to the outlet exit. 

\begin{figure}[H]
    \centering
        \includegraphics[width=\textwidth]{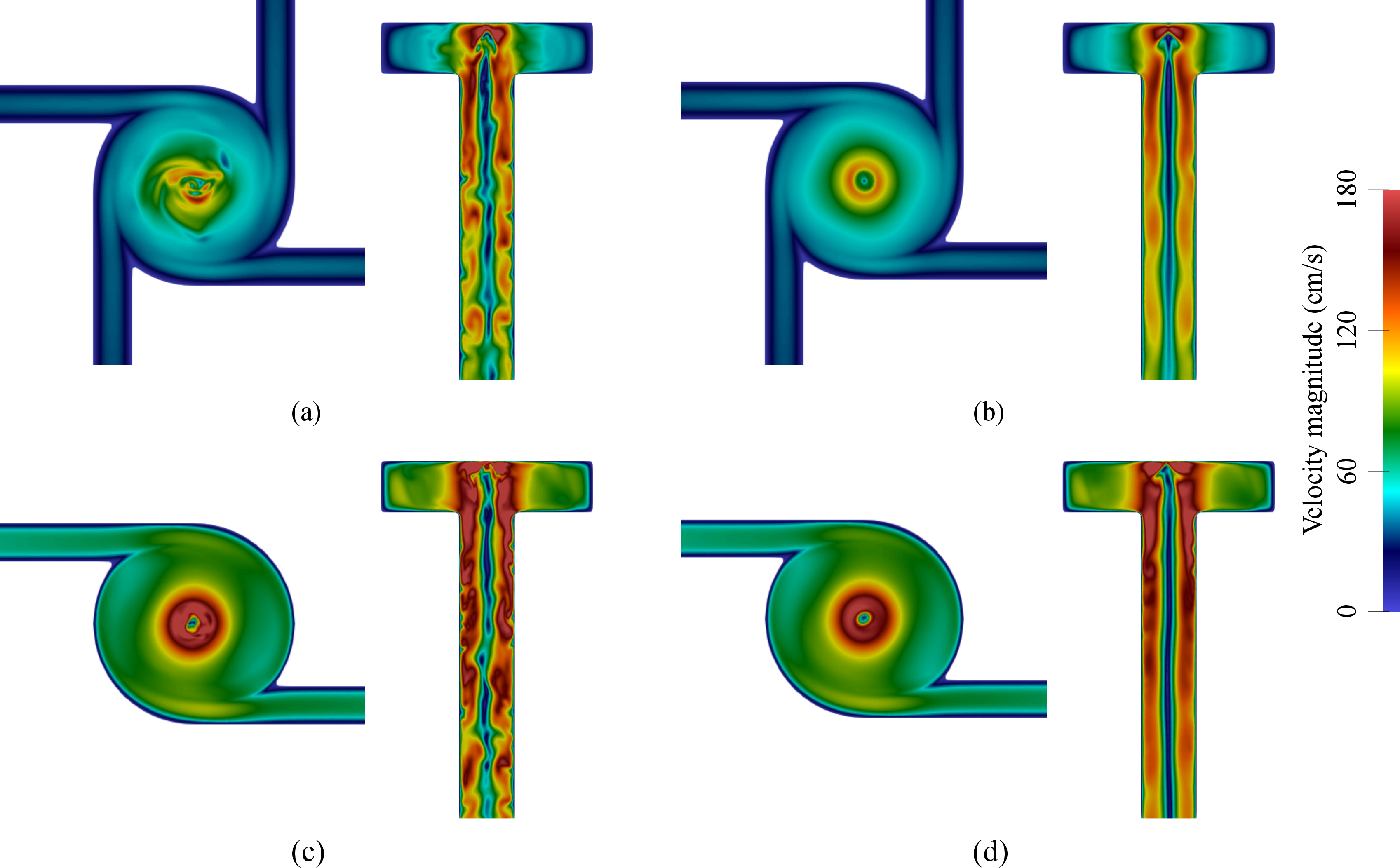}
\caption{Comparison of velocity magnitude for four-inlet (top row) and two-inlet (bottom row) MIVM, both operating at 80 mL/min total flow rate. (a) and (c) are instantaneous snapshots, and (b) and (d) are averaged over time.
}
    \label{fig:velocity_mivm}
\end{figure}

Figure~\ref{fig:velocity_mivm} shows the velocity magnitude contours for the four-inlet (top row) and two-inlet (bottom row) MIVM mixers. 
In the top views, the two-inlet mixer shows higher velocity magnitudes.
This is as expected since each inlet carries twice the flow of those in the four-inlet mixer.
Moreover, the fluid in the mixing chamber of the four-inlet mixer accelerates more towards the outlet due to the lower inlet flow rate for each channel.
In the outlet channel, we can observe that the velocity magnitude is higher for the two-inlet mixer.
Since the total flow rate in the mixers is identical, the higher magnitude indicates a large angular velocity in the two-inlet outlet. 
In other words, the two-inlet mixer exhibits a strong swirling flow in the outlet channel, whereas the four-inlet case displays weaker swirling and a more even outlet flow profile.
The swirling flow also affects the flow near the centerline of the outlet, where the center of fluid rotation creates a stagnation zone.
Due to stronger swirling motion, the velocity magnitude in the two-inlet mixer remains close to zero over a longer distance in the outlet channel compared to that of the four-inlet mixer.

\begin{figure}[H]
    \centering
    \includegraphics[width=\textwidth]{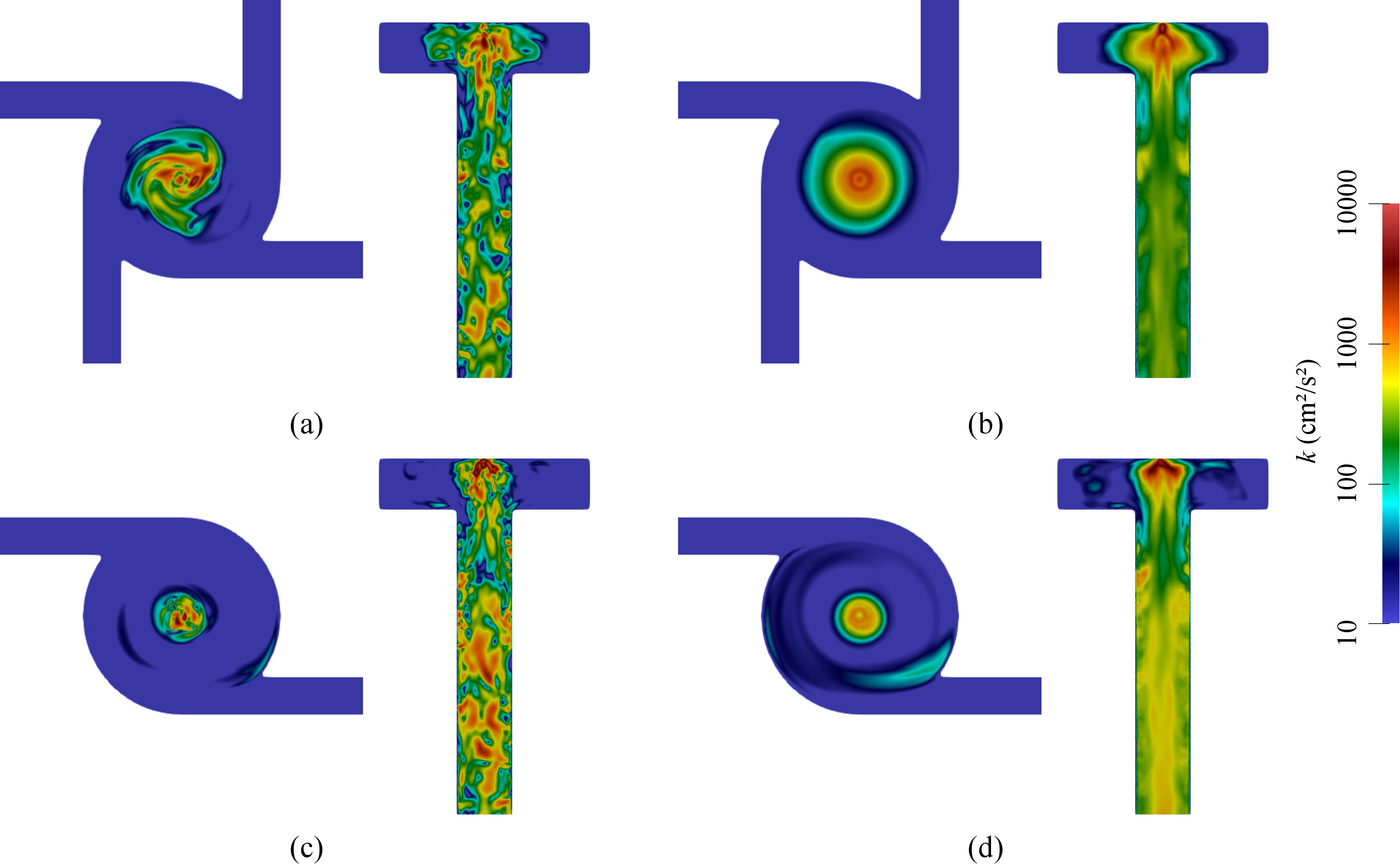}
\caption{Comparison of turbulent kinetic energy, $k$, for four-inlet (top row) and two-inlet (bottom row) MIVM, both operating at 80 mL/min total flow rate. (a) and (c) are instantaneous snapshots, and (b) and (d) are averaged over time. Note that the color scale is logarithmic. }
    \label{fig:tke_mivm}
\end{figure}

Since the mixing performance is heavily dependent on the turbulence characteristics of the flow, we calculate the turbulent kinetic energy, $k$, as
\begin{equation}
    k = \frac{1}{2} \lVert \bl u' \rVert^2,
    \label{eqn:tke}
\end{equation}
where $\bl u'$ is the velocity fluctuation vector, calculated by subtracting the time-averaged velocity vector, $\bar{\bl u}$, from the velocity vector, $\bl u$, as $\bl u' = \bl u -\bar{\bl u}$.  
Figure~\ref{fig:tke_mivm} shows the turbulent kinetic energy for the two-inlet and four-inlet MIVM. 
The four-inlet mixer exhibits stronger turbulence inside the mixing chamber with non-zero values of turbulent kinetic energy over a larger portion of the mixing chamber than the two-inlet mixer.
In contrast, the two-inlet mixer has close to zero turbulent kinetic energy in the mixing chamber, except near the center where the fluid exits to the outlet channel. 
This indicates late turbulence development in the mixing chamber of the two-inlet MIVM.
Once the flow enters the outlet channel, higher turbulent kinetic energy levels are observed for the two-inlet case, potentially caused by the stronger swirling motion. 
For both MIVM configurations, the turbulent kinetic energy is well sustained downstream of the outlet channel. 

\begin{figure}[h!]
    \centering
        \includegraphics[width=0.47\textwidth]{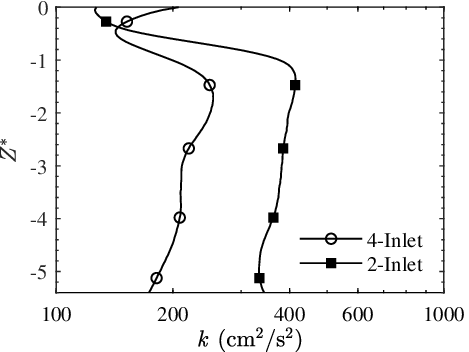}
\caption{Average turbulent kinetic energy, $k$, along the non-dimensionalized MIVM outlet channel axial direction, $Z^*$. The turbulent kinetic energy is averaged first over the cross-section along $Z^*$ at each time step, and then time-averaged over all time steps.}
    \label{fig:tke_plot_mivm}
\end{figure}

To help illustrate the turbulence evolution in the outlet channel, Figure~\ref{fig:tke_plot_mivm} shows the time and cross-sectional averaged turbulent kinetic energy, $\bar{k}$, along the outlet channel. 
The four-inlet mixer maintained a consistent level of turbulent kinetic energy, around $200 \;\mathrm{cm^2/s^2}$, along the outlet channel. 
However, the two-inlet mixer experienced a period of rapid turbulence development at the beginning of the outlet channel, where the turbulent kinetic energy increases from around $150 \;\mathrm{cm^2/s^2}$ to $400 \;\mathrm{cm^2/s^2}$. 

\begin{figure}[H]
    \centering
        \includegraphics[width=0.8\textwidth]{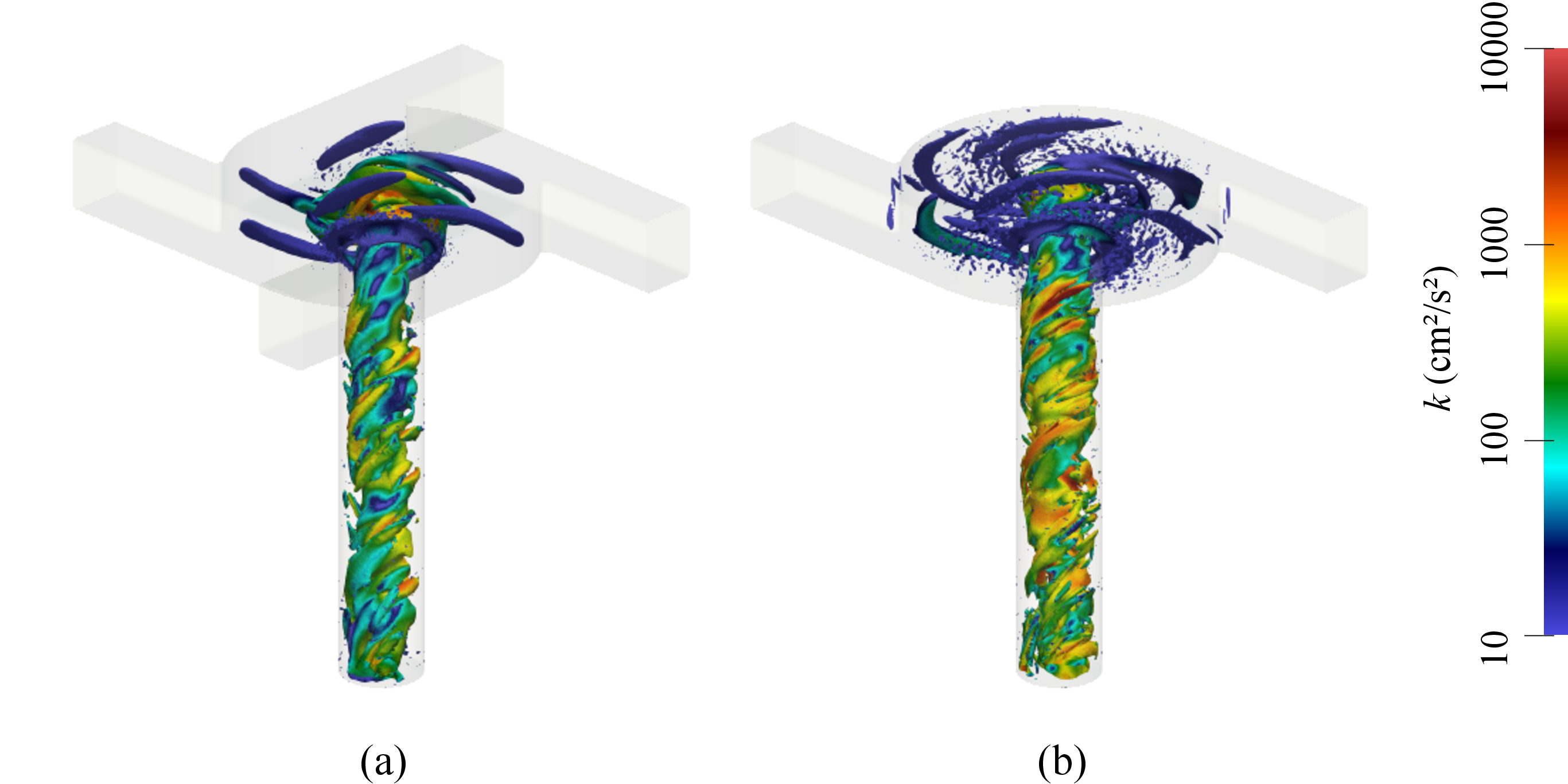}
\caption{Instantaneous q-criterion isosurfaces ($Q = 2\times 10^5$) for (a) four-inlet MIVM and (b) two-inlet MIVM. The isosurfaces are colored with turbulent kinetic energy, $k$.}
    \label{fig:mivm_q_tke}
\end{figure}

Figure~\ref{fig:mivm_q_tke} provides instantaneous snapshots of turbulence vortex structures in the two-inlet and four-inlet MIVM using Q-criterion isosurfaces, colored by turbulent kinetic energy value.
In both mixers, a tube-shaped turbulent structure fills the outlet channel, with swirling micro-structures. 
For the four-inlet mixer, the tip of the vortex tube expands into the mixing chamber, indicating turbulence onset in the mixing chamber.
The two-inlet case, on the other hand, presents little turbulent structure in the mixing chamber. 
It is also important to highlight that the swirling structure is more horizontal in the two-inlet mixer than in the four-inlet mixer.
This distinction indicates a stronger centrifugal turbulent flow in the two-inlet mixer outlet. 

\subsection{Confined impinging jets mixer}

\begin{figure}[H]
    \centering
        \includegraphics[width=0.8\textwidth]{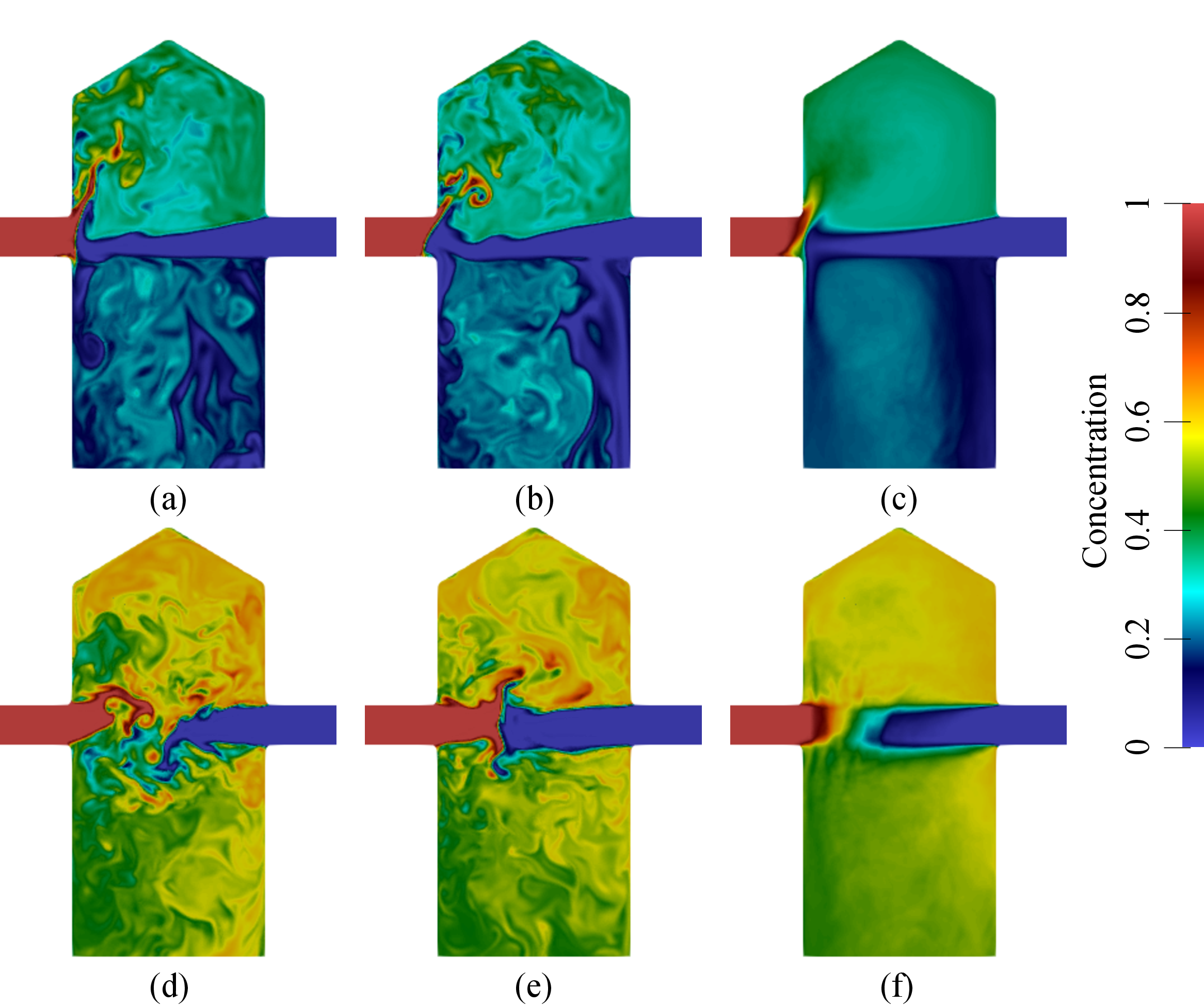}
 \caption{Comparison of concentration fields in the CIJM for flow-rate ratios of 1:3 (top row) and 3:3 (bottom row). (a), (b), (d), and (e) are instantaneous snapshots, and (c) and (f) are averaged over time. All plots are taken at the center plane of the mixer.}
    \label{fig:concentraiton_cij}
\end{figure}

Figure~\ref{fig:concentraiton_cij} shows the concentration contours for the CIJM operating at two different ethonal-to-water flow rate ratios (FRRs), 1:3 and 3:3. 
Note that the flow rate is kept constant for the water stream at 60 mL/min, while the ethanol flow rates are 20 mL/min and 60 mL/min for FFR 1:3 and 3:3, respectively. 
In the 1:3 case, there is a strong momentum imbalance of the two streams due to the flow rate difference, as well as a lower density of ethanol compared to water.
As a result, the ethanol stream is pinned at its inlet to the mixing chamber by the water stream. 
Despite mixing occurring to some capacity inside the mixing chamber, this extreme offsetting of the impingement point inside the mixing chamber results in a significant volume of water exiting the mixer unmixed. 
As evident in Figure~\ref{fig:concentraiton_cij}c, the bottom right corner of the mixer appears in dark blue, which corresponds to almost pure water. 

In contrast, the FRR 3:3 case has equal volumetric flow rates for ethanol and water, resulting in an approximately central impinging point. 
Nonetheless, the lower density and higher viscosity of ethanol compared to water cause the ethanol stream to carry less momentum than the water stream entering the mixing chamber. 
As a result, the impingement point is offset slightly to the side of the ethanol inlet.
An important new observation is that the fluid aggregating at the tip of the mixing chamber contains more ethanol (concentration $\approx 0.6$, yellow color).
Note that this is not due to any buoyancy effect since gravity is not modeled in this study. 
This behavior is likely caused by the combination of inlet jets asymmetry and geometry asymmetry between the top and bottom of the mixing chamber.
This newfound phenomenon could have a significant effect on the final product, such as nanoparticles. 

\begin{figure}[H]
    \centering
    \begin{subfigure}[b]{0.47\textwidth}
        \includegraphics[width=\textwidth]{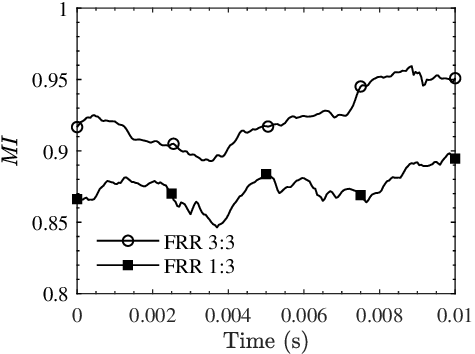}
        \caption{ }
    \end{subfigure}
    \hfill
    \begin{subfigure}[b]{0.47\textwidth}
        \includegraphics[width=\textwidth]{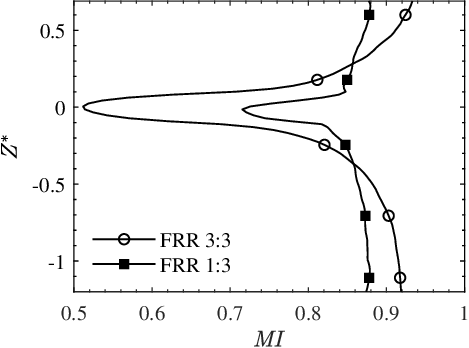}
        \caption{ }
    \end{subfigure}
   \caption{Comparison of mixing index for the CIJM operating under FFRs 1:3 and 3:3. (a) Temporal evolution of the mixing index ($MI$) at the outlet of the mixer. (b) Average mixing index along the mixing chamber vertical direction, $Z^*$.}
    \label{fig:MIcomparison_cij}
\end{figure}

Similar to the MIVM results, to quantify the mixing performance, the mixing index is calculated from the simulation results. 
Figure~\ref{fig:MIcomparison_cij}a shows the temporal evolution of the mixing index (MI) at the outlet of the mixing chamber. 
The CIJM operating at FRR~3:3 maintains mixing index values above 0.9. 
In contrast, the FRR~1:3 case produces lower mixing index values between 0.85 and 0.9, indicating a more non-uniform mixture at the outlet.
Figure~\ref{fig:MIcomparison_cij}b shows the mixing index of each cross-section along the vertical direction of the mixing chamber.
Under both operating conditions, the mixing index exhibits a rapid improvement near the impinging plane of the two streams ($Z^*=0)$.

\begin{figure}[H]
    \centering
    \includegraphics[width=0.8\textwidth]{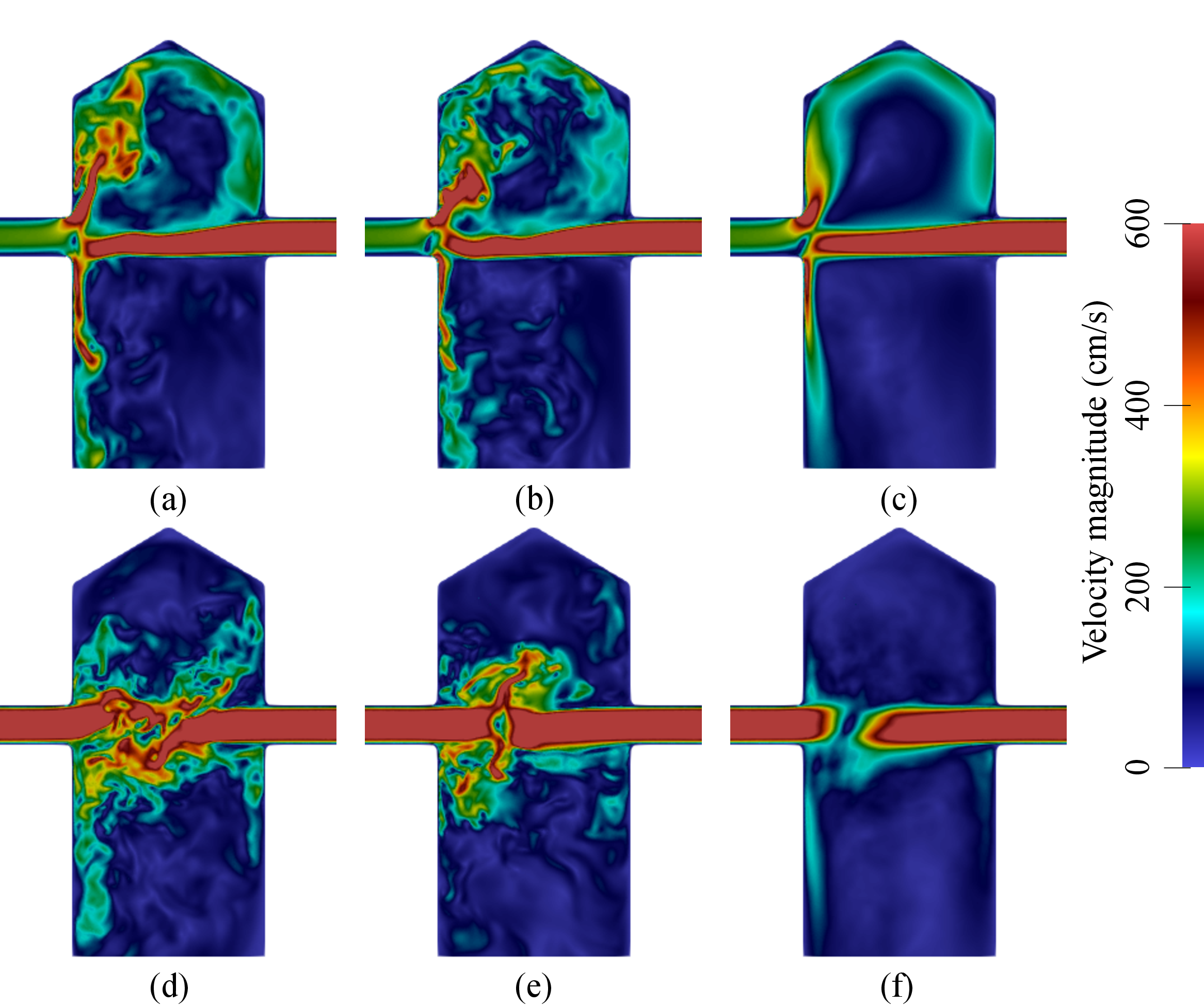}
\caption{Comparison of velocity magnitude in the CIJM for flow-rate ratios of 1:3 (top row) and 3:3 (bottom row). (a), (b), (d), and (e) are instantaneous snapshots, and (c) and (f) are averaged over time. All plots are taken at the center plane of the mixer.}
    \label{fig:velocity_cij}
\end{figure}

The velocity magnitude contours at the center plane are shown in Figure~\ref{fig:velocity_cij}.  
Due to the flow imbalance in the FRR~1:3 case, the ethanol stream follows a circular motion in the top region of the mixing chamber.
This weakens the turbulence intensity since the fluid momentum is not fully transformed into turbulent kinetic energy upon impingement. 
On the other hand, as shown in Figure~\ref{fig:velocity_cij}f, under FFR 3:3, the velocity magnitude significantly diminishes from the point of impingement, signaling the flow streams' kinetic energy converting to turbulence energy. 

\begin{figure}[H]
    \centering
    \includegraphics[width=0.8\textwidth]{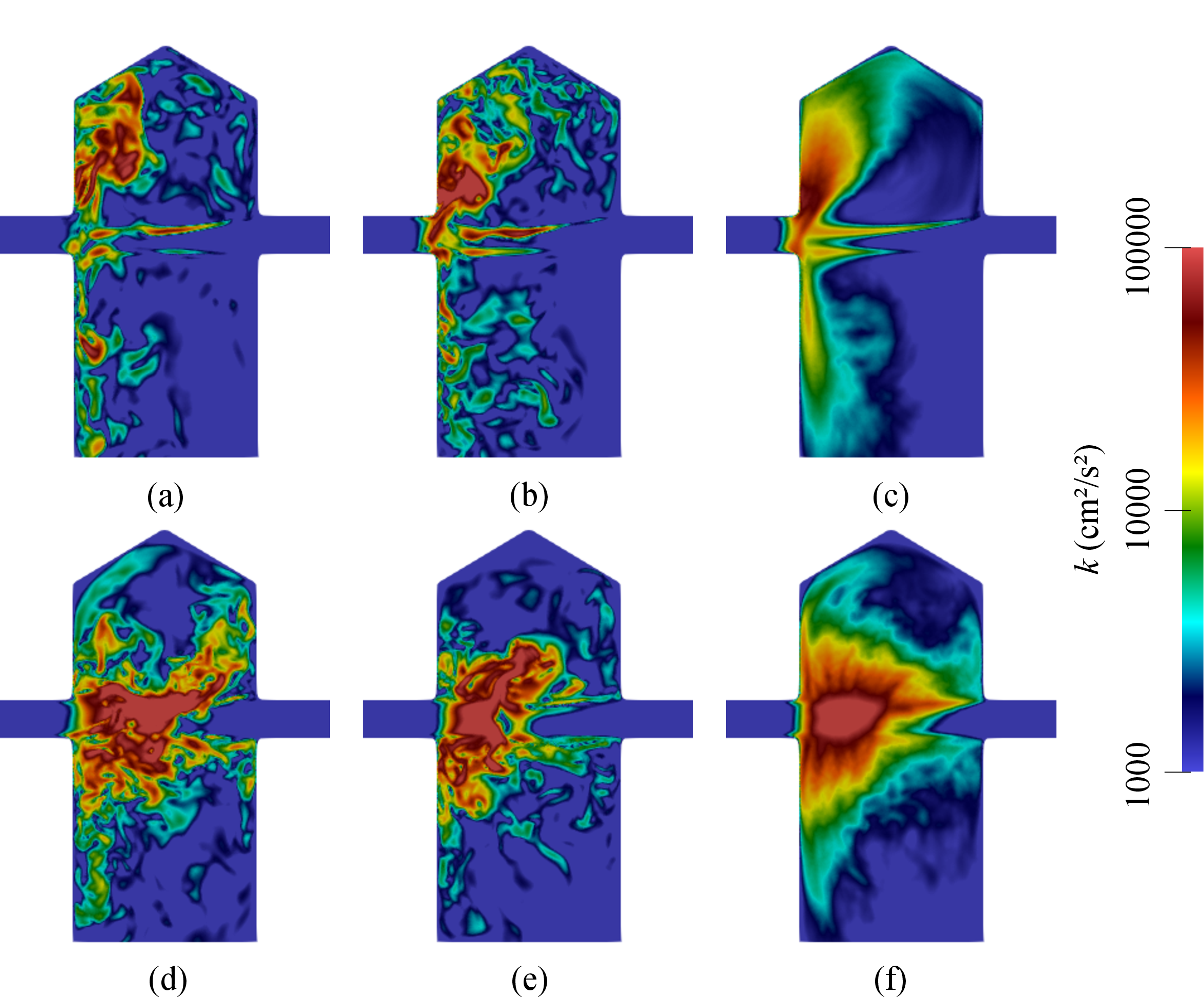}
    \caption{Comparison of turbulent kinetic energy, $k$, fields in the CIJM for flow-rate ratios of 1:3 (top row) and 3:3 (bottom row). (a), (b), (d), and (e) are instantaneous snapshots, and (c) and (f) are averaged over time. All plots are taken at the center plane of the mixer. Note that the legend color is on a logarithmic scale.}
    \label{fig:tke_cij}
\end{figure}

This conversion of energy can be further visualized with the turbulent kinetic energy contours in Figure~\ref{fig:tke_cij}. 
As can be seen from the instantaneous and time-averaged results, under the FRR~3:3 condition (Figure~\ref{fig:tke_cij}d-f), high levels of turbulent kinetic energy are generated at the center of the mixing chamber originating from the impingement point.
Note that the color is on a logarithmic scale. 
The turbulence expands uniformly outwards while rapidly decaying. 
In contrast, as shown in Figure~\ref{fig:tke_cij}a-c, under the imbalance FFR 1:3 condition, the turbulence generates primarily as a result of the ethonal stream interacting with the mixing chamber wall as it is pinned against the wall by the water stream. 
This results in a "wake-like" wall turbulence structure. 
Moreover, we can see two lines of intense turbulence in the center of the mixing chamber.
This is generated by the high-velocity water stream jetting into low-velocity fluid in the chamber, creating a hollow-cylindrical shear turbulence layer.
Neither of these two modes of turbulence, boundary layer and shear, is the intended method of turbulence generation for CIJM through impingement. 
Furthermore, since the turbulence mixing does not originate from the impingement point, the fresh water and ethanol streams are primarily interacting with already partially-mixed fluids inside the mixing chamber, which can be an undesirable feature in terms of final product quality. 

\begin{figure}[H]
    \centering
    \includegraphics[width=0.47\textwidth]{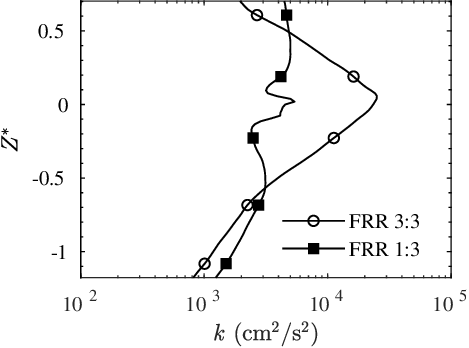}
\caption{Average turbulent kinetic energy, $k$, along the non-dimensionalized CIJM mixing chamber axial direction, $Z^*$. The turbulent kinetic energy is averaged first over the cross-section along $Z^*$ at each time step, and then time-averaged over all time steps.}
    \label{fig:tke_cij_plot}
\end{figure}

Figure~\ref{fig:tke_cij_plot} shows the time-averaged and cross-sectional-averaged turbulence kinetic energy, $k$, along the vertical direction of the mixing chamber, $Z^*$, non-dimensionalized by the chamber diameter.
At the impinging plane, the FRR~3:3 case has a high turbulent kinetic energy level of $k \approx2.5\times 10^4$ cm\textsuperscript{2}/s\textsuperscript{2}. 
Away from the impingement plane, the turbulent energy exhibits an approximately $-3/2$ power law decay.
On the other hand, the different boundary-layer turbulence for the FFR 1:3 case results in an almost constant level of turbulent kinetic energy along the vertical direction.
The maximum level of turbulent energy is around $5\times 10^3$ cm\textsuperscript{2}/s\textsuperscript{2}.

\begin{figure}[H]
    \centering
        \includegraphics[width=0.8\textwidth]{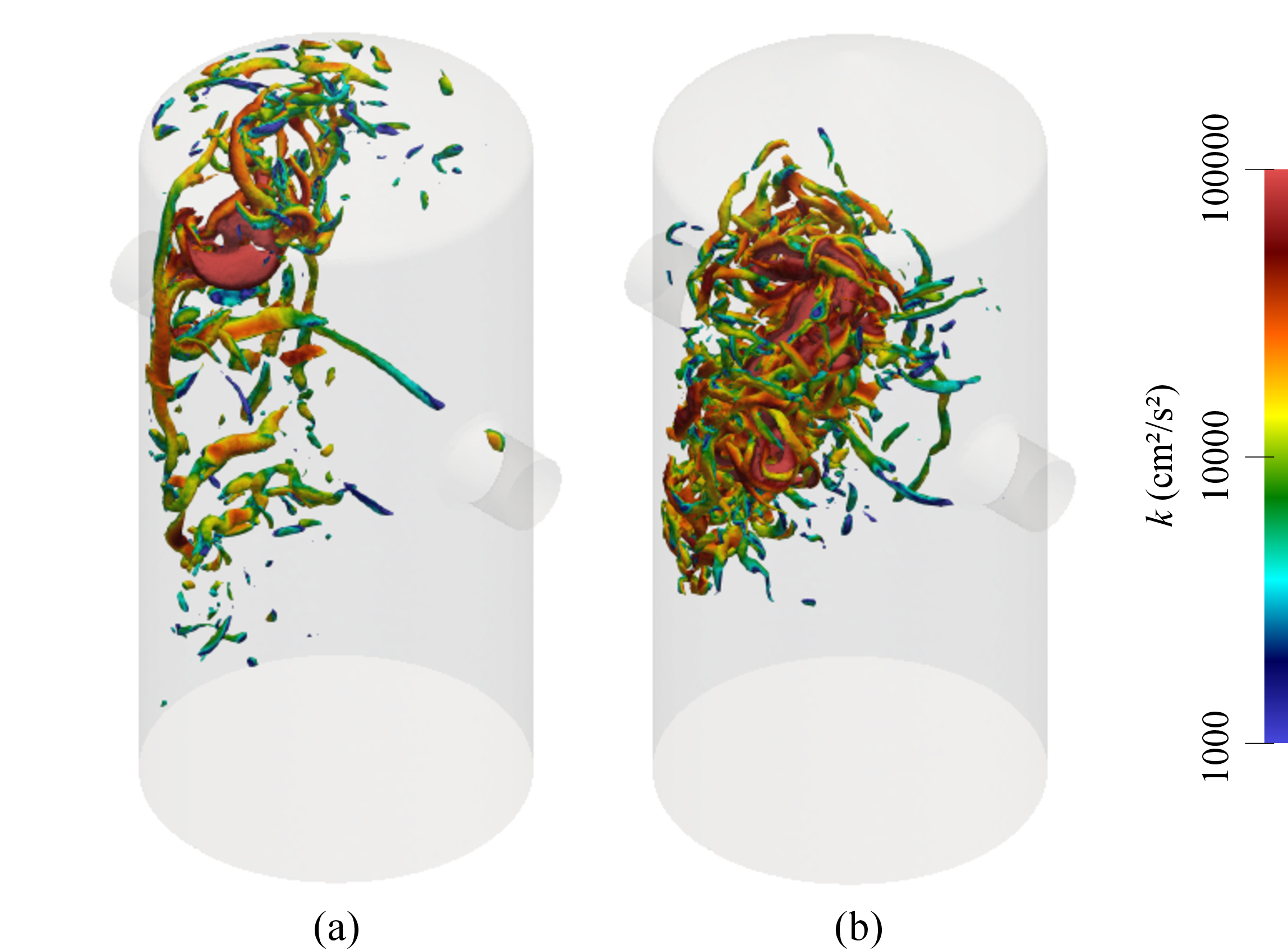}
\caption{Instantaneous Q-criterion isosurfaces ($Q = 5\times 10^8$) for the CIJM operating at FRR 1:3 (a) and FRR 3:3 (b), colored by the turbulent kinetic energy,$k$.}
    \label{fig:tke_q_cij}
\end{figure}

Figure~\ref{fig:tke_q_cij} shows instantaneous Q-criterion isosurfaces colored by the turbulent kinetic energy, $k$, for both flow rate ratios. 
Corroborating with the previous observations, the turbulence in the FFR 1:3 case primarily exists near the wall region on the side of the ethanol stream inlet.
Whereas the turbulence in the FRR~3:3 case originates from the center impingement point and carries a spherical shape. 

\section{Discussion}

As demonstrated by the results, the RBVMS finite element method produces high-fidelity CFD solutions that predict turbulence structures in fine detail.
For the MIVM, we can clearly see the breakup of the streams and onset of turbulence inside the mixing chamber, which is not observed in previously published RANS or DES simulations.
Qualitatively, it is clear that the solutions are on par with DNS by comparing the complex turbulent ring structures around the impingement point of the CIJM~\cite{hao2020flow}, while our study added an extra layer of complexity with non-uniform density and viscosity. 
These characteristics of the simulation results stem from the consistent formulation of the subgrid scale turbulent modeling of the RBVMS method, which is superior in accuracy and resolution to those of commercially available software in this specific application. 

These favorable results obtained from the RBVMS finite element method come with its unique challenges related to the numerical characteristics. 
One major challenge is the solution of the advection-diffusion equation.
Physically, when the two flow streams first come into contact, it creates an interfacial plane. 
In a numerical sense, the concentration field is close to discontinuous at this fluid interface.
Capturing this interface is challenging since the Galerkin method used to discretize the solution space relies on the assumption that the solution is continuous. 
This challenge is further exacerbated by the low diffusivity of ethanol in water, which makes the problem advection-dominant, therefore requiring a strong stabilization scheme. 
This challenge is addressed through the combination of three techniques described in the methods section: a physics-based mesh refinement scheme, an additional constant in the stabilization parameter, and a bounded approach in calculating mixture density and viscosity. 

For the MIVM, the novelty of this study lies in comparing the two-inlet and four-inlet designs. 
The turbulence and mixing dynamics of the four-inlet MIVM operating at similar Reynolds numbers have been reported in existing literature~\cite{shi2013measurements,liu2015batchelor,liu2015flow}. 
Specifically, the mean velocity contour (Figure~\ref{fig:velocity_mivm}b) closely resembles a result in a previously published experimental paper~\cite{shi2013measurements}, with a similar peak velocity value, the same ring-shaped high velocity zone, and a low velocity center. 
The zero-velocity zone along the centerline of the outlet channel, or ``backmixing", has also been reported and studied in a recent publication~\cite{peng2024study}.
In this study, we observed that the two-inlet design exacerbates backmixing by generating stronger spiral motion in the outlet channel, creating a larger zero-velocity zone in its center. 
The turbulent mixing behavior inside the mixing chamber shows a layered circulating motion that transitions into a forced-vortex region towards the center (Figure~\ref{fig:concentration_mivm}a), in line with published experimental observations.
The transition is earlier in the two-inlet mixer due to the higher Reynolds number, as predicted by literature~\cite{liu2017turbulent}. 
On the other hand, the four-inlet MIVM exhibits higher turbulent kinetic energy in the mixing chamber than in the two-inlet configuration, whereas turbulence is stronger in the outlet channel of the two-inlet MIVM than that of the four-inlet MIVM. 
These characteristics can affect product quality in industrial applications of these mixers, such as in the manufacture of nanoparticles, where specimens in the four-inlet MIVM experience weaker turbulence for a longer duration, whereas specimens in the two-inlet MIVM experience stronger turbulence for a shorter duration.

The operating condition of the CIJM in this study distinguishes itself from existing literature in two major aspects: (i) the Reynolds number is over 2000 in this study, whereas the existing studies are mostly limited to values under 1000~\cite{tucker1980mixing,fonte2015flow}. 
Despite differences in Reynolds number, we observed similar turbulent and mixing behavior.
For the 1:1 flow rate ratio, we observed the formation of a shear layer at the impinging point~\cite{wood1991experimental,li2007modelling} (Figure~\ref{fig:concentraiton_cij}e) and the helical instabilities around impinging jets ~\cite{hao2020flow} (Figure~\ref{fig:tke_q_cij}b).
At the Reynolds number simulated, we also observed the breaking and reappearance of the shear layer at the impinging point, as observed in a previous experimental study for $\mathrm{Re}>500$~\cite{li2014experimental} (Figure~\ref{fig:concentraiton_cij}d and~\ref{fig:concentraiton_cij}e). 
For the imbalanced flow rate ratio of 1:3, we observed a similar shift in the impinging point and a reduction in mixing quality as in previous experiments~\cite{fonte2015flow}.
In addition, we present a new discovery of the weaker jet being pinned at the inlet and creating a wake-like recirculation zone at the top of the mixing chamber (Figure~\ref{fig:velocity_cij}a-c). 
(ii) In this study, the two inlet jets carry different fluids, resulting in non-uniform density and viscosity, which adds complexity to the fluid dynamics. 
When the flow rates of the two inlets are identical, the averaged jet impingement point is off-center in the mixing chamber.
The location of the impingement point is offset to the side of ethanol due to two effects: the lower density of ethanol, resulting in less kinetic energy, and the higher viscosity of ethanol that dissipates energy faster.
Based on the established observation that an offset in the impinging point in the CIJM will always result in a decrease in the mixing quality~\cite{fonte2015flow}, we can theoretically improve the mixing performance by adjusting the total energy of the two streams to control the impingement position. 
This can be achieved by controlling the mean velocity of each stream, which can be adjusted by changing either the flow rate ratio or the inlet tube diameter ratio.

Comparing the two types of mixers, the turbulence characteristics and their relation to mixing are drastically different.
The onset of turbulence is almost instantaneous from the flow impingement in the CIJM, whereas in the MIVM, the turbulence exhibits a more gradual development.
This difference in turbulence generation results in a significant difference in turbulent kinetic energy.
The maximum value in the CIJM is two to three orders of magnitude larger than that of the MIVM.
On the other hand, the turbulent kinetic energy in the CIJM decays with a power law away from the jet impingement plane, which is significantly faster than that of the MIVM, which turbulent kinetic energy is sustained through the outlet channel. 
Another distinction is the turbulence relation with the mixing performance.
For the CIJM, the turbulence is the strongest at the point of impingement, where the two fluids are at their initial unmixed state.
As for the MIVM, with the more gradual development of turbulence, the fluid is already partially mixed when the flow reaches peak turbulence level, near the mixing chamber outlet.
This difference can have implications for applications such as nanoparticle production, where the mixing time scale, turbulence time scale, and particle assembly time scale all need to be in alignment. 

Lastly, we would like to acknowledge the limitations of this study. 
The simulation results of this study do not include quantitative experimental validation under the exact same operating conditions. 
This is due to the lack of existing literature on ethanol-water mixtures operating at Reynolds numbers similar to those in this study, especially for the CIJM. 
An in-house experimental validation of the results will require significant additional resources and is out of the scope of this study.
However, we would like to highlight the extensive numerical work undertaken in this study to reduce numerical error and ensure the validity of the results, as well as the qualitative similarity of the results to the existing experimental and computational literature. 
Another limitation arises from the RBVMS method adapted in this study, which fundamentally models the subgrid velocity of the flow field.
Therefore, the turbulence microstructure or micro-mixing behaviors will not be available below the grid scale. 
To predict nanoparticle reaction and formation using macro and meso-scale turbulence information, various modeling approaches have been proposed in the literature~\cite{schwarzer2006predictive,bagheri2019population,shin2025mechanistic}.
Nonetheless, coupling these models with the stabilized finite element method remains a challenge.

\section{Conclusion}
In this study, we simulated the fluid and mixing dynamics of the multi-inlet vortex mixer (MIVM) and the confined impinging jets mixer (CIJM) using the residual-based variational multiscale finite element method. 
We demonstrated that the chosen simulation method produces high turbulence fidelity with reasonable computational cost. 
The computational framework used in this study should be considered the standard for similar mixer simulation applications.
We compared the performance of two-inlet and four-inlet MIVM designs and showed that the fluid is better mixed for the four-inlet configuration. 
For the CIJM, we showed the impact of the inflow ratio of the two streams on the fluid dynamics and mixing performance. 
In a comparison of the two mixers, the MIVM generates turbulence through shear, while the CIJM triggers turbulence through flow impingement.
The turbulent kinetic energy is sustained through the outlet channel of the MIVM.
In contrast, the turbulent energy peaks at the impingement point in the CIJM and exhibits a power law decay towards the outlet.
The maximum turbulent kinetic energy in the CIJM is around two orders of magnitude higher than that in the MIVM. 

\section*{Data availability}
The simulation files and results presented in this study are available on request from the corresponding authors.

\section*{Acknowledgment}

The authors would like to acknowledge that this study is supported by Eli Lilly and Company. The authors would like to thank Michael Robert Tullis for providing comments to the original draft. 

\section*{Author contributions statement}
Conceptualization, D.J., M.M., K.R., and A.A.; methodology, D.J., M.M., and A.A.; software, D.J. and M.M.; validation, D.J.; formal analysis, D.J. and M.M.; investigation, D.J.; resources, A.A.; data curation, D.J. and M.M.; writing---original draft preparation, D.J. and M.M.; writing---review and editing, D.J., K.R., and A.A.; visualization, D.J. and M.M.; supervision, D.J., K.R., and A.A.; project administration, K.R. and A.A. All authors have read and agreed to the published version of the manuscript.

\bibliographystyle{unsrt}
\bibliography{ref}

\appendix
\appendixpage
\addappheadtotoc
\linespread{1}

\section{Mass conservation verification}
\label{app:mass}
\begin{figure}[H]
    \centering
    \begin{subfigure}[b]{0.47\textwidth}
        \includegraphics[width=\textwidth]{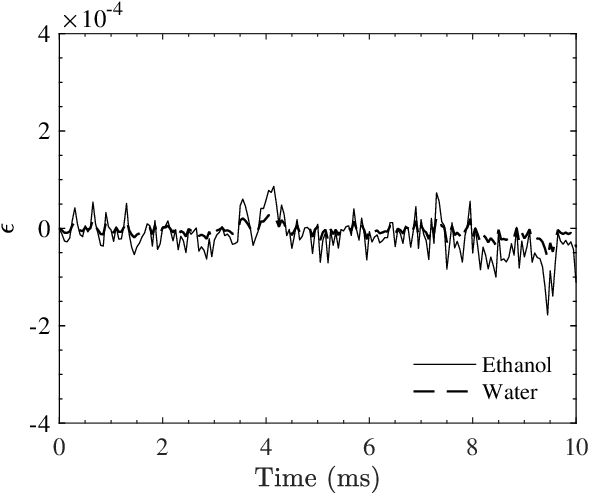}
        \caption{ }
    \end{subfigure}
    \hfill
    \begin{subfigure}[b]{0.47\textwidth}
        \includegraphics[width=\textwidth]{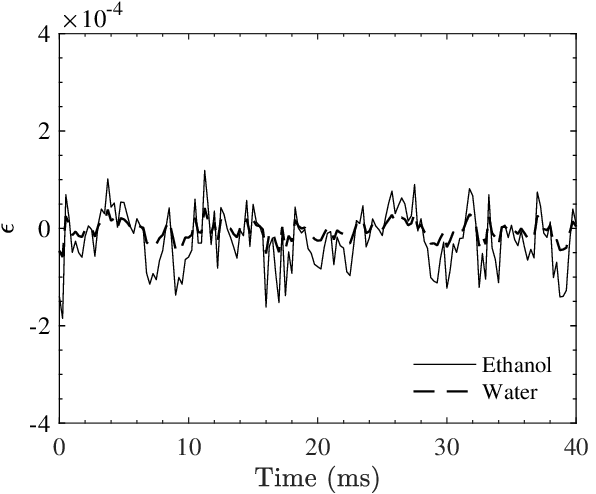}
        \caption{ }
    \end{subfigure}
   \caption{Numerical error in mass conservation for (a) CIJM operating at 1:1 flow rate and (b) four inlet MIVM. }
    \label{fig:mass_error}
\end{figure}

In this study, we used a clipping approach to eliminate the numerical overshoot or undershoot of the concentration scaler when calculating the mixed fluid density and viscosity, defined in~\eqref{eqn:rho_mu_bound}. 
Owing to the high mesh resolution used in this study, the amount of overshoot or undershot is small and only affects a limited number of nodes. 
Therefore, this implementation has minimal effect on the conservation of mass in the fluid domain. 
To demonstrate this effect quantitatively, we define the numerical error in total mass change, as
\begin{equation}
    e_m = \int\frac{\partial \rho}{\partial t} \dd \Omega + \int \rho \bl u\cdot \bl n \dd \Gamma,
    \label{eqn:mass}
\end{equation}
where $\Omega$ and $\Gamma$ are the fluid domain and domain boundary, respectively, and $\bl n$ is the normal direction at the boundary.
On the right hand side of~\eqref{eqn:mass}, the first term represents the change of total mass in the volume, and the second term represents the total mass flux at the boundary.
To non-dimensionality this error, we define the numerical error in total mass as
\begin{equation}
    \epsilon := \frac{e_m\Delta t}{m},
    \label{eqn:masserror}
\end{equation}
where $\Delta t$ is the time step size, and $m$ is the mass in the domain. 
$\epsilon$ represents the added/subtracted artificial mass due to the numerical scheme at each time step.
The error in mass conservation is shown in Figure~\ref{fig:mass_error} for both water and ethanol for two of the cases ran for this study.
The numerical mass conservation error is well below $0.1\%$ at all times. 
Furthermore, all errors oscillate around 0, meaning that no significant amount of numerical mass error is induced over time. 

\end{document}